\documentclass[aps,prd,twocolumn,floatfix,superscriptaddress,preprintnumbers,nofootinbib]{revtex4}

\usepackage{slashed}
\usepackage{amsmath,amssymb,graphicx} 
\usepackage{epsf,color}
\usepackage{hyperref}
\usepackage{cancel}
\usepackage{framed}

\definecolor{nicered}{rgb}{0.5,0.,0.}
\definecolor{nicegreen}{rgb}{0.,0.5,0.}
\definecolor{niceblue}{rgb}{0.,0.,0.5}
\hypersetup{colorlinks,citecolor=nicegreen,linkcolor=nicered,urlcolor=niceblue}

\begin{document}
\title{Collider Signatures of Flavorful Higgs Bosons}
\date{October 7, 2016}

\def\Cincy{Department of Physics, University of Cincinnati, Cincinnati, Ohio 45221, USA}
\def\Fermilab{Fermi National Accelerator Laboratory, P.O. Box 500, Batavia, IL 60510, USA}

\author{Wolfgang~Altmannshofer}
\email[Electronic address:]{altmanwg@ucmail.uc.edu}
\affiliation{\Cincy}

\author{Joshua~Eby}
\email[Electronic address:]{ebyja@mail.uc.edu}
\affiliation{\Cincy}
\affiliation{\Fermilab}

\author{Stefania~Gori}
\email[Electronic address:]{stefania.gori@ucmail.uc.edu}
\affiliation{\Cincy}

\author{Matteo~Lotito}
\email[Electronic address:]{lotitomo@mail.uc.edu}
\affiliation{\Cincy}

\author{Mario~Martone}
\email[Electronic address:]{martonmo@ucmail.uc.edu}
\affiliation{\Cincy}

\author{Douglas~Tuckler}
\email[Electronic address:]{tuckleds@mail.uc.edu}
\affiliation{\Cincy}

\begin{abstract}
Motivated by our limited  knowledge of the Higgs couplings to first two generation fermions, we analyze the collider phenomenology of a class of two Higgs doublet models (2HDMs) with a non-standard Yukawa sector. One Higgs doublet is mainly responsible for the masses of the weak gauge bosons and the third generation fermions, while the second Higgs doublet provides mass for the lighter fermion generations.  
The characteristic collider signatures of this setup differ significantly from well-studied 2HDMs with natural flavor conservation, flavor alignment, or minimal flavor violation. New production mechanisms for the heavy scalar, pseudoscalar, and charged Higgs involving second generation quarks can become dominant. The most interesting decay modes include $H/A \to c c, tc, \mu\mu, \tau\mu$ and $H^\pm \to cb, cs, \mu\nu$. 
Searches for low mass di-muon resonances are currently among the best probes of the heavy Higgs bosons in this setup.
\end{abstract}

\maketitle

\allowdisplaybreaks

\section{Introduction} \label{sec:intro}

The LHC measurements of Higgs rates~\cite{Khachatryan:2014jba,Aad:2015gba,Khachatryan:2016vau} show an overall good agreement with Standard Model (SM) predictions. By now it is established that the couplings of the Higgs to the weak gauge bosons are SM-like to a good approximation. This implies that the main origin of the weak gauge bosons' mass is the vacuum expectation value (vev) of the 125~GeV Higgs boson. Also the masses of the top quark, the bottom quark and the tau lepton appear to be largely due to the 125~GeV Higgs, as indicated by the measured values of Higgs couplings to the third generation fermions.

However, little is known about the origin of the masses of the first and second generation fermions. Direct measurements of the Higgs couplings to these fermions are challenging. In the lepton sector, ATLAS and CMS will eventually reach sensitivity to the Higgs coupling to muons at the SM level by measuring the $h \to \mu^+\mu^-$ decay rate~\cite{CMS:2013xfa}. The Higgs coupling to electrons is tiny and the $h \to e^+ e^-$ rate in the SM is far beyond the experimental reach of the LHC. Sensitivities to the Higgs electron coupling not far above the SM might be reached at future $e^+ e^-$ colliders running on the Higgs pole~\cite{Altmannshofer:2015qra,d'Enterria:2016yqx}. In the quark sector, various ideas have been explored to determine the coupling of the Higgs to charm quarks. Those include the measurement of the exclusive $h \to J/\psi \gamma$ decay rate~\cite{Bodwin:2013gca,Aad:2015sda,Perez:2015lra,Koenig:2015pha}, inclusive $h \to c \bar c$ measurements using 
charm-tagging techniques~\cite{Delaunay:2013pja,Perez:2015aoa,Perez:2015lra}, and Higgs production in association with charm quarks~\cite{Brivio:2015fxa}.
The rates of the exclusive Higgs decays $h \to \phi \gamma$, $h \to \rho \gamma$, and $h \to \omega \gamma$ are sensitive to the Higgs couplings to strange, down, and up quarks~\cite{Kagan:2014ila,Aaboud:2016Zpg}.
Also the Higgs $p_T$ distribution~\cite{Bishara:2016jga,Soreq:2016rae,Bonner:2016sdg} and the $W^\pm h$ charge asymmetry~\cite{Yu:2016rvv} have sensitivity to the light quark couplings.

While inclusive $h \to c \bar c$ measurements might reach SM sensitivities at a future 100~TeV collider~\cite{Perez:2015lra} and will be quite precisely determined at future $e^+e^-$ colliders \cite{Asner:2013psa}, the Higgs couplings to strange, down, and up quarks remain out of direct experimental reach in the foreseeable future, unless they are enhanced by orders of magnitude, if compared to SM expectations.  

Given the limited sensitivities of the direct measurements of the Higgs couplings to the light generations, we are led to develop complementary strategies to identify the origin of the masses of first and second generation.
In this work we study the possibility that the origin of the first and second generation fermion masses is {\it not} the 125~GeV Higgs but an additional source of electro-weak symmetry breaking as proposed in~\cite{Altmannshofer:2015esa} (see also~\cite{Ghosh:2015gpa,Botella:2016krk,Das:1995df,Blechman:2010}) and study the implications. 
Arguably the simplest realization of such a setup is a two Higgs doublet model (2HDM) where one doublet (that we approximately identify as the 125~GeV Higgs) couples mainly to the third generation, while a second doublet couples mainly to the first and second generation.
One motivation, with regards to fermion mass generation, is a reduction of the Yukawa coupling hierarchy between the third and the lighter generations via a Higgs vev hierarchy.

In such a framework we expect distinct phenomenological implications at low and high energy experiments.
A generic prediction are flavor-violating couplings of the 125~GeV Higgs~\cite{Altmannshofer:2015esa,Ghosh:2015gpa,Botella:2016krk} which could explain the small hint for the lepton flavor-violating Higgs decay $h \to \tau \mu$ at CMS~\cite{Khachatryan:2015kon}. Other signatures include rare lepton flavor-violating $B$ meson decays like $B \to K^{(*)} \tau\mu$ with branching ratios as large as $10^{-7}$ and the rare top decay $t \to c h$ with branching ratios as large as $10^{-3}$~\cite{Altmannshofer:2015esa}.

In this paper we determine the characteristic collider signatures of the second Higgs doublet.
We find that novel production mechanisms involving second generation quarks can become dominant for moderate and large $\tan\beta$. The largest production mode of the neutral Higgs bosons is production from a $c \bar c$ initial state. The charged Higgs bosons are dominantly produced from a $c s$ initial state.
The most interesting decay modes include $H/A \to cc, tc, \mu\mu, \tau\mu$ and $H^\pm \to cb, cs,  \mu\nu$.
Our work provides continued motivation to search for low mass di-muon resonances and low mass di-jet resonances. Searches for di-muon resonances are currently the best probes of the considered setup, while searches for di-jet resonances have sensitivities similar to the ``traditional'' di-tau searches for additional neutral Higgs bosons. 

The paper is organized as follows. 
In Sec.~\ref{sec:2HDM} we discuss aspects of the proposed 2HDM framework that are relevant for our analysis, focusing in particular on the couplings of the heavy Higgs bosons. In Sec.~\ref{sec:SMhiggs} the modifications to the properties of the $125$~GeV Higgs are analysed and confronted with Higgs coupling measurements at the LHC. In Sec.~\ref{sec:neutralhiggs}, we discuss the collider phenomenology of the heavy neutral Higgs bosons and identify distinct features in production and decay modes. The production and decay modes of the charged Higgs are discussed in Sec.~\ref{sec:chargedhiggs}. In Sec.~\ref{sec:signals} we discuss the constraints that can be derived using current searches for heavy Higgs bosons and show predictions for novel collider signatures.
We conclude in Sec.~\ref{sec:conclusions}.

\section{Two Flavorful Higgs Doublets} \label{sec:2HDM}

The considered setup is a 2HDM in which one Higgs doublet is mainly responsible for the mass of the third generation of SM fermions, while the second Higgs doublet gives masses mainly to the first and second generations. 
We start by briefly reviewing generic 2HDMs (see e.g.~\cite{Branco:2011iw,Crivellin:2013wna}) in Sec.~\ref{sec:review}. In Sec.~\ref{sec:rank1} we discuss the specific Yukawa textures of our model and the resulting heavy Higgs couplings.

\subsection{Generic Two Higgs Doublet Models} \label{sec:review}

The two Higgs doublets with hypercharge $+1/2$ are denoted $\Phi$ and $\Phi'$ and decompose as
\begin{equation}
 \Phi = \begin{pmatrix} \phi^+ \\ \frac{1}{\sqrt{2}} ( v + \phi + i a) \end{pmatrix}  \, , ~ \Phi^\prime = \begin{pmatrix} \phi^{\prime +} \\ \frac{1}{\sqrt{2}} ( v^\prime + \phi^\prime + i a^\prime) \end{pmatrix} \, ,
\end{equation}
where $v^2 + v^{\prime 2} = v_W^2 = (246$~GeV$)^2$ is the SM Higgs vacuum expectation value (vev) squared and the ratio of Higgs vevs is $\tan\beta = t_\beta = v/v^\prime$. Note that in generic two Higgs doublet models, the Higgs fields $\Phi$ and $\Phi^\prime$ can be transformed into each other, and the ratio of Higgs vevs is therefore a basis dependent quantity~\cite{Davidson:2005cw}.

For simplicity we will not consider CP violation in the Higgs sector.\footnote{Note that the Yukawa couplings of the Higgs bosons to quarks necessarily contain complex parameters to reproduce the phase in the CKM matrix and will lead to CP violation in the Higgs potential at the loop level. However, such effects are generically small and will be neglected here.}
In this case, after electroweak symmetry breaking, the components of $\Phi$ and $\Phi^\prime$ mix in the following way to form mass eigenstates
\begin{eqnarray}
\begin{pmatrix} \phi^+ \\ \phi^{\prime +} \end{pmatrix} &=& \begin{pmatrix} s_\beta & - c_\beta \\ c_\beta & s_\beta \end{pmatrix} \begin{pmatrix} G^+ \\ H^+ \end{pmatrix} \, , \\
\begin{pmatrix} a \\ a^\prime \end{pmatrix} &=& \begin{pmatrix} s_\beta & - c_\beta \\ c_\beta & s_\beta \end{pmatrix} \begin{pmatrix} G^0 \\ A \end{pmatrix} \, , \\
\begin{pmatrix} \phi \\ \phi^\prime \end{pmatrix} &=& \begin{pmatrix} c_\alpha & s_\alpha \\ -s_\alpha & c_\alpha \end{pmatrix} \begin{pmatrix} h \\ H \end{pmatrix} \, ,
\end{eqnarray}
with $c_x \equiv \cos x$, $s_x \equiv \sin x$ for $x = \alpha, \beta$. 
The three states $G^0$, $G^\pm$ provide the longitudinal components of the $Z$ and $W^\pm$ gauge bosons. The remaining physical states consist of two CP-even scalars $h$ and $H$,
one CP-odd scalar $A$, and the charged Higgs $H^\pm$. We will identify $h$ with the SM-like Higgs with a mass of $m_h \simeq 125$~GeV. The heavy Higgs bosons $H$, $A$, and $H^\pm$ are approximately degenerate in the decoupling limit, $m_H \simeq m_A \simeq m_{H^\pm}$, with mass splittings of $\mathcal{O}(v_W^2/m_A^2)$.
In the decoupling limit, the mixing angle $\alpha$ is strongly related to $\beta$ 
with $\alpha = \beta - \frac{\pi}{2} + \mathcal{O}(v_W^2/m_A^2)$.

Turning to the interactions of the two Higgs doublets with the SM fermions, the most general Yukawa Lagrangian can be written as
\begin{eqnarray}
 -\mathcal L_Y &=& \sum_{i,j} \Big( \lambda^u_{ij} (\bar q_i u_j) \tilde\Phi + \lambda^{\prime u}_{ij} (\bar q_i u_j) \tilde \Phi' \nonumber \\
 && ~~~~+ \lambda^d_{ij} (\bar q_i d_j) \Phi + \lambda^{\prime d}_{ij} (\bar q_i d_j) \Phi' \nonumber \\
 && ~~~~+ \lambda^e_{ij} (\bar\ell_i e_j) \Phi + \lambda^{\prime \ell}_{ij} (\bar\ell_i e_j) \Phi' \Big)~ + \text{h.c.} ~, \label{LY}
\end{eqnarray}
where $\tilde \Phi^{(\prime)} = i \sigma_2 (\Phi^{(\prime)})^*$.
The $q_i$, $\ell_i$ are the three generations of left-handed quark and lepton doublets, and the $u_i$, $d_i$, $e_i$ are the right-handed up quark, down quark, and charged lepton singlets. (A discussion of neutrino masses and mixing is beyond the scope of this work.)
In all generality, the mass matrices of the charged SM fermions receive contributions from both Higgs doublets.
In the fermion mass eigenstate basis we use the notation
\begin{eqnarray}
 m^{(\prime) u}_{q q^\prime} = \frac{v^{(\prime)}}{\sqrt{2}} \langle q_L |\mathcal \lambda^{(\prime) u} | q_R^\prime \rangle ~,&& \quad \text{for}~ q,q^\prime \in \{u,c,t\} ~, \\
 m^{(\prime) d}_{q q^\prime} = \frac{v^{(\prime)}}{\sqrt{2}} \langle q_L |\mathcal \lambda^{(\prime) d} | q_R^\prime \rangle ~,&& \quad \text{for}~ q,q^\prime \in \{d,s,b\} ~, \\\label{mll}
 m^{(\prime)}_{\ell \ell^\prime} = \frac{v^{(\prime)}}{\sqrt{2}} \langle \ell_L |\mathcal \lambda^{(\prime) \ell} | \ell_R^\prime \rangle ~,&& \quad \text{for}~ \ell,\ell^\prime \in \{e,\mu,\tau\} ~.
\end{eqnarray}
Notice that in the mass basis the matrices $m^{(\prime)}_{xx^\prime}$, with $x=q,\ell$, are not diagonal. The couplings of the physical neutral Higgs bosons to the fermions can be parameterized as
\begin{eqnarray}
 \mathcal L &\supset& - \sum_{i,j} (\bar u_i P_R u_j) \left( h (Y_h^u)_{ij} + H (Y_H^u)_{ij} + i A (Y_A^u)_{ij} \right) \nonumber \\
 && - \sum_{i,j} (\bar d_i P_R d_j) \left( h (Y_h^d)_{ij} + H (Y_H^d)_{ij} + i A (Y_A^d)_{ij} \right) \nonumber \\
 && - \sum_{i,j} (\bar \ell_i P_R \ell_j) \left( h (Y_h^\ell)_{ij} + H (Y_H^\ell)_{ij} + i A (Y_A^\ell)_{ij} \right) \nonumber \\
 && + \text{h.c.} ~.
\end{eqnarray}
In the fermion mass eigenstate basis we find for the flavor-diagonal Higgs couplings
\begin{eqnarray}\label{exlept1}
 Y_\ell^h &\equiv& \langle \ell_L | Y_h^\ell| \ell_R \rangle = \frac{m_\ell}{v_W} \left( \frac{c_\alpha}{s_\beta} - \frac{m_{\ell\ell}^\prime}{m_\ell} \frac{c_{\beta - \alpha}}{s_\beta c_\beta} \right)  ~, \\\label{exlept2}
 Y_\ell^H  &\equiv& \langle \ell_L | Y_H^\ell| \ell_R \rangle = \frac{m_\ell}{v_W} \left( \frac{s_\alpha}{s_\beta} + \frac{m_{\ell\ell}^\prime}{m_\ell} \frac{s_{\beta - \alpha}}{s_\beta c_\beta} \right) ~, \\ \label{exlept3}
 Y_\ell^A  &\equiv& \langle \ell_L | Y_A^\ell| \ell_R \rangle = \frac{m_\ell}{v_W} \left( \frac{1}{t_\beta} - \frac{m_{\ell\ell}^\prime}{m_\ell} \frac{1}{s_\beta c_\beta} \right) ~, 
\end{eqnarray}
for $\ell = e, \mu, \tau$, and analogous for the quark couplings. $m_\ell$ are the mass eigenvalues that is, from \eqref{mll}, $m_\ell=m_{\ell\ell}+m^\prime_{\ell\ell}$. For the flavor-violating Higgs couplings we obtain
\begin{eqnarray} \label{eq:yllp}
 Y_{\ell \ell^\prime}^h  &\equiv& \langle \ell_L | Y_h^\ell| \ell_R^\prime \rangle = -\frac{m_{\ell\ell^\prime}^\prime}{v_W}\frac{c_{\beta - \alpha}}{s_\beta c_\beta} ~, \\
 Y_{\ell \ell^\prime}^H  &\equiv& \langle \ell_L | Y_H^\ell| \ell_R^\prime \rangle = +\frac{m_{\ell\ell^\prime}^\prime}{v_W}  \frac{s_{\beta - \alpha}}{s_\beta c_\beta} ~, \\ 
 Y_{\ell \ell^\prime}^A  &\equiv& \langle \ell_L | Y_A^\ell| \ell_R^\prime \rangle = -\frac{m_{\ell\ell^\prime}^\prime}{v_W}  \frac{1}{s_\beta c_\beta} ~, 
\end{eqnarray}
for $\ell \neq \ell^\prime$ and $\ell, \ell^\prime = e, \mu, \tau$. Analogous expressions hold for the flavor-violating quark couplings.

We write the couplings of the charged Higgs bosons to the fermions as
\begin{eqnarray}\label{eq:chargedL}
 \mathcal L &\supset& - \sqrt{2} \sum_{i,j} \Big(  (\bar d_i P_R u_j)  H^- (Y_\pm^u)_{ij} + (\bar u_i P_R d_j)  H^+ (Y_\pm^d)_{ij} \nonumber \\
 && ~~~~~ + (\bar \nu_i P_R \ell_j)  H^+ (Y_\pm^\ell)_{ij}  \Big) + \text{h.c.} ~.
\end{eqnarray}
In the fermion mass eigenstate basis we find for the couplings to quarks
\begin{eqnarray}
 Y_{qq^\prime}^\pm  &\equiv& \langle q_L | Y_\pm^d| q_R^\prime \rangle \nonumber \\
 &=& \frac{m_{q^\prime}}{v_W} \left( \frac{V_{qq^\prime}}{t_\beta} - \sum_{x=d,s,b} \frac{m_{xq^\prime}^\prime}{m_{q^\prime}} \frac{V_{qx}}{s_\beta c_\beta} \right)~, 
\end{eqnarray}
for $q \in \{u,c,t\}$ and $q^\prime \in \{d,s,b\}$, and 
\begin{eqnarray}
 Y_{qq^\prime}^\pm  &\equiv& \langle q_L | Y_\pm^u| q_R^\prime \rangle \nonumber \\
 &=& \frac{m_{q^\prime}}{v_W} \left( \frac{V_{q^\prime q}^*}{t_\beta} - \sum_{x=u,c,t} \frac{m_{xq^\prime}^\prime}{m_{q^\prime}} \frac{V_{xq}^*}{s_\beta c_\beta} \right) ~, 
\end{eqnarray}
for $q \in \{d,s,b\}$ and $q^\prime \in \{u,c,t\}$. In the above expressions, $V$ is the CKM matrix.
In the lepton sector, we neglect neutrino mixing as it is of no relevance for our considerations. We find
\begin{equation}\label{eq:chargedLept}
 Y_{\ell}^\pm  \equiv \langle \nu_\ell | Y_\pm^\ell| \ell_R \rangle = \frac{m_\ell}{v_W} \left( \frac{1}{t_\beta} - \frac{m_{\ell\ell}^\prime}{m_\ell} \frac{1}{s_\beta c_\beta} \right) ~.
\end{equation}
The physical couplings of the Higgs bosons to the fermions are completely determined by the two angles $\alpha$ and $\beta$, the mass matrices $m^\prime$ in the fermion mass eigenstate basis, and the known masses of the SM fermions, as well as CKM elements.
Note that none of the expressions above assumes a specific Yukawa texture. The expressions hold in any 2HDM.

\subsection{Yukawa Textures} \label{sec:rank1}

Our setup imposes that the Yukawa couplings of $\Phi$ are rank 1 and that they provide mass only to one generation of fermions, that will become the dominant component of the third generation.
This assumption singles out a Higgs basis and renders the ratio of Higgs vevs, $\tan\beta$, well defined and physical.

We start with a discussion of the lepton sector.
In the flavor basis where the $\Phi$ lepton Yukawa is diagonal, we consider the following Yukawa texture
\begin{equation}
 \lambda^\ell \sim \frac{\sqrt{2}}{v} \begin{pmatrix} 0 & 0 & 0 \\ 0 & 0 & 0 \\ 0 & 0 & m_\tau \end{pmatrix} \,,~
 \lambda^{\prime \ell} \sim \frac{\sqrt{2}}{v^\prime} \begin{pmatrix} m_e & m_e & m_e \\ m_e & m_\mu & m_\mu \\ m_e & m_\mu & m_\mu \end{pmatrix} \,. \label{eq:l_texture}
\end{equation}
This texture gives the observed lepton masses, and it can naturally explain the hierarchy between second and third generation lepton masses, if $v^\prime\ll v$, or equivalently $\tan\beta\gg 1$.
Next we rotate into the mass eigenstate basis. Expanding in $m_\mu/m_\tau$ and $m_e/m_\mu$ we find
\begin{eqnarray}
 m_{ee}^\prime &=& m_e + \mathcal{O}(m_e^2/m_\tau) \,, \\ 
 m_{\mu\mu}^\prime &=& m_\mu + \mathcal{O}(m_\mu^2/m_\tau) \,,\label{mmu} \\
 m_{\tau\tau}^\prime &=& \mathcal{O}(m_\mu) ~\,, \\
 m_{e\mu}^\prime &=& \mathcal{O}(m_e m_\mu /m_\tau) \,,~ m_{\mu e}^\prime = \mathcal{O}(m_e m_\mu /m_\tau) \,,  \\
 m_{e\tau}^\prime &=& \mathcal{O}(m_e) \,,~~~~~~~~~~~ m_{\tau e}^\prime = \mathcal{O}(m_e) \,, \\
 m_{\mu\tau}^\prime &=& \mathcal{O}(m_\mu) \,,~~~~~~~~~~~  m_{\tau\mu}^\prime = \mathcal{O}(m_\mu) \,. \label{eq:mtaumu}  
\end{eqnarray}
The diagonal entries for the first and second generation are to a good approximation determined by the corresponding observed lepton masses. Note that $e - \mu$ mixing is not given by an entry of $\mathcal{O}(m_e)$ as one could na\"ively expect, but is additionally suppressed. The reason for this suppression is a global $U(2)_\ell \times U(2)_e$ flavor symmetry acting on the first two generation of leptons that is only broken by a single Yukawa coupling $\lambda^{\prime \ell}$.
This suppression of $e - \mu$ mixing is sufficient to avoid the stringent constraints from flavor-violating low energy transitions like $\mu \to e \gamma$ or $\mu \to e$ conversion~\cite{preparation}.

It seems natural to assume analogous Yukawa textures also in the quark sector. However, in addition to reproducing quark masses, the quark Yukawas also have to conform with the observed values of the CKM quark mixing matrix. Given that the hierarchies in the down-quark masses and the hierarchies among CKM elements are comparable, while the hierarchies in the up-quark masses are considerably larger, we will assume that the quark mixing is mainly generated from the down Yukawas.

The up sector can then be chosen in a way completely analogous to the lepton sector.
The expressions (\ref{eq:l_texture})-(\ref{eq:mtaumu}) hold with the replacements $e \to u$, $\mu \to c$, and $\tau \to t$. The strongly suppressed $u -c$ mixing guarantees that constraints from neutral $D$ meson oscillations are easily avoided in this setup~\cite{preparation}.

A down-quark Yukawa texture that naturally leads to the observed down-quark masses and CKM mixing angles reads
\begin{equation}
 \lambda^d \sim \frac{\sqrt{2}}{v} \begin{pmatrix} 0 & 0 & 0 \\ 0 & 0 & 0 \\ 0 & 0 & m_b \end{pmatrix} \,,~
 \lambda^{\prime d} \sim \frac{\sqrt{2}}{v^\prime} \begin{pmatrix} m_d & \lambda m_s & \lambda^3 m_b \\ m_d & m_s & \lambda^2 m_b \\ m_d & m_s & m_s \end{pmatrix} \,, \label{eq:d_texture}
\end{equation}
with the Cabibbo angle $\lambda \simeq 0.23$.
To a reasonable approximation one has $\lambda^2 m_b \sim m_s$, while $\lambda^3 m_b$ and $\lambda m_s$ are a factor of few larger than $m_d$. We consider this small missmatch acceptable at the level of Yukawa textures. 

Rotating into the mass eigenstate basis we find
\begin{eqnarray}
 m_{dd}^\prime &=& m_d + \mathcal{O}(m_s \lambda^4) \,, \\ 
 m_{ss}^\prime &=& m_s + \mathcal{O}(m_s \lambda^2) \,, \\
 m_{bb}^\prime &=& \mathcal{O}(m_s) ~\,, \\
 m_{ds}^\prime &=& \mathcal{O}(m_s \lambda^3) \,,~~\, m_{sd}^\prime = \mathcal{O}(m_d \lambda^2) \,,  \\
 m_{db}^\prime &\simeq& -m_b V_{td}^* \,,~~~~ m_{bd}^\prime = \mathcal{O}(m_d) \,, \\
 m_{sb}^\prime &\simeq& -m_b V_{ts}^* \,,~~~~  m_{bs}^\prime = \mathcal{O}(m_s) \,.  
\end{eqnarray}
Note that the $m_{db}^\prime$ and $m_{sb}^\prime$ parameters are approximately fixed by the requirement to quantitatively reproduce the CKM mixing matix.
The fact that $d - s$ mixing is suppressed, and at most of order $m_s \lambda^3$, eases constraints from neutral Kaon oscillations. Nevertheless, Kaon, $B_d$, and $B_s$ meson oscillations do put constraints on the size of the $m_{sd}^\prime$ $m_{bd}^\prime$, and $m_{bs}^\prime$ parameters and on $\tan\beta$ depending on the heavy Higgs masses~\cite{preparation}.
The flavor-violating entries in the down sector have only a relevant impact on the production of the heavy Higgses in association with b-quarks (see Sec. \ref{sec:neutralhiggs} below). In order to avoid the constraints from meson oscillations, one could use the following $\lambda^{\prime d}$ Yukawa coupling
\begin{equation}
 \lambda^{\prime d} \simeq \frac{\sqrt{2}}{v^\prime} \begin{pmatrix} m_d & \lambda m_s & \lambda^3 m_b \\ 0 & m_s & \lambda^2 m_b \\ 0 & 0 & m_s \end{pmatrix} \,,
\end{equation}
which would lead to a production cross section of the heavy Higgses in association with b-quarks that is approximately a factor of 2 smaller compared to the texture in Eq.~(\ref{eq:d_texture}).
In the following we will stick to the texture in Eq.~(\ref{eq:d_texture}), keeping in mind that meson mixing might constrain the production of the heavy Higgses in association with b-quarks to be as much as a factor 2 smaller than what shown in the plots of Fig.~\ref{fig:sigma}.

We now combine the Yukawa textures specified above with the generic expressions for heavy Higgs couplings discussed in Sec.~\ref{sec:review}. 
For the flavor-diagonal heavy Higgs couplings to taus we find
\begin{eqnarray} \label{eq:kappa_tau_H}
 \kappa_\tau^H = \frac{Y_\tau^H}{Y_\tau^\text{SM}} &=& \frac{1}{t_\beta} \frac{s_\alpha}{c_\beta} + \mathcal{O}\left(\frac{m_\mu}{m_\tau}\right)\times \frac{t_\beta}{s_\beta^2} s_{\beta-\alpha} \,, \\
 \kappa_\tau^A = \frac{Y_\tau^A}{Y_\tau^\text{SM}} &=& \frac{1}{t_\beta} + \mathcal{O}\left(\frac{m_\mu}{m_\tau}\right)\times \frac{t_\beta}{s_\beta^2} \,, \\
 \kappa_\tau^\pm = \frac{Y_{\nu_\tau \tau}^\pm}{Y_\tau^\text{SM}} &=& \frac{1}{t_\beta} + \mathcal{O}\left(\frac{m_\mu}{m_\tau}\right)\times \frac{t_\beta}{s_\beta^2} \,,
\end{eqnarray}
and analogous expressions hold for the couplings to third generation quarks. The leading terms in these couplings are suppressed for moderate and large $\tan\beta$ with respect to the SM Higgs couplings.
The corrections that are suppressed by the ratio of second to third generation masses are proportional to $\tan\beta$ and can actually dominate in the large $\tan\beta$ regime. 

For the couplings to muons, the second term in \eqref{exlept1}-\eqref{exlept3} is no longer sub-dominant. From \eqref{mmu}, $m^\prime_{\mu\mu}/m_{\mu}=1+\mathcal{O}(m_{\mu}/m_{\tau})$, so we find
\begin{eqnarray}
 \kappa_\mu^H = \frac{Y_\mu^H}{Y_\mu^\text{SM}} &=& t_\beta \frac{c_\alpha}{s_\beta} + \mathcal{O}\left(\frac{m_\mu}{m_\tau}\right)\times \frac{t_\beta}{s_\beta^2} s_{\beta-\alpha} \,, \\
 \kappa_\mu^A = \frac{Y_\mu^A}{Y_\mu^\text{SM}} &=& -t_\beta + \mathcal{O}\left(\frac{m_\mu}{m_\tau}\right)\times \frac{t_\beta}{s_\beta^2} \,, \\
 \kappa_\mu^\pm = \frac{Y_{\nu_\mu \mu}^\pm}{Y_\mu^\text{SM}} &=& -t_\beta + \mathcal{O}\left(\frac{m_\mu}{m_\tau}\right)\times \frac{t_\beta}{s_\beta^2} \,.
\end{eqnarray}
Analogous expressions hold for the second generation quark couplings and for the couplings to first generation fermions.
Note the enhancement of these couplings by $\tan\beta$.
Flavor off-diagonal couplings of the heavy Higgses between second and third generation are generically of the same order as the corresponding flavor-diagonal couplings to the second generation.
In the lepton sector we have for example
\begin{eqnarray}
 \kappa_{\mu \tau}^H = \frac{Y_{\mu \tau}^H}{Y_\tau^\text{SM}} &=& \mathcal{O}\left(\frac{m_\mu}{m_\tau}\right)\times \frac{t_\beta}{s_\beta^2}  s_{\beta-\alpha} \,, \\
 \kappa_{\mu \tau}^A = \frac{Y_{\mu \tau}^A}{Y_\tau^\text{SM}} &=& \mathcal{O}\left(\frac{m_\mu}{m_\tau}\right)\times \frac{t_\beta}{s_\beta^2} \,.
\end{eqnarray}
Analogous expressions hold for the flavor-violating couplings involving the second and third generation of quarks.
Flavor-violating couplings of the charged Higgs to quarks contain additional terms that are proportional to small CKM elements. For example
\begin{equation}
  \kappa_{st}^\pm = \frac{Y_{s t}^\pm}{Y_t^\text{SM}} \simeq \frac{V_{ts}}{t_\beta} + \mathcal{O}\left(\frac{m_c}{m_t}\right) \times \frac{t_\beta}{s_\beta^2} \,.
\end{equation}

Given these couplings, the collider phenomenology of the heavy Higgs bosons in our model can be markedly different, if compared to less flavorful 2HDM setups that have been studied extensively in the literature~\cite{Alves:2012ez, Craig:2012vn, Altmannshofer:2012ar, Bai:2012ex, Grinstein:2013npa, Chen:2013rba, Craig:2013hca, Barger:2013ofa, Carena:2013ooa, Chang:2013ona, Cheung:2013rva, Celis:2013ixa, Ferreira:2014naa, Kanemura:2014bqa, Craig:2015jba}.\footnote{See also~\cite{Arroyo:2013tna,Botella:2015hoa,Buschmann:2016uzg,Sher:2016rhh,Cvetic:1997zd,Cvetic:1998uw,Bauer:2015kzy,Bauer:2015fxa} for  studies of interesting 2HDM setups with new sources of flavor violation.}
In contrast to models with natural flavor conservation~\cite{Glashow:1976nt}, flavor alignment~\cite{Pich:2009sp,Altmannshofer:2012ar} or minimal flavor violation~\cite{D'Ambrosio:2002ex,Buras:2010mh}, the couplings of the heavy Higgses to fermions are not proportional to the fermion masses.
For moderate and large values of $\tan\beta$, the couplings to the third generation fermions are suppressed, while the couplings to the second and first generation are enhanced, if compared to the couplings of the SM Higgs.
Therefore, the branching ratios do not have to be dominated by decays to third generation (top, bottom, tau), and we expect sizable branching ratios involving charm quarks and muons. Moreover, new non-standard production modes for the heavy Higgs bosons involving light quark generations can become relevant.

Before discussing the corresponding heavy Higgs collider phenomenology in detail in Secs.~\ref{sec:neutralhiggs},~\ref{sec:chargedhiggs}, and~\ref{sec:signals}, we briefly outline the modified properties of the 125~GeV Higgs and the implied constraints on the parameter space.

\section{Properties of the SM-like Higgs} \label{sec:SMhiggs}

In our model, the couplings of the light Higgs to SM particles are generically modified.
The existing measurements of the Higgs rates at the LHC depend crucially on the Higgs couplings to vector bosons and to the third generation fermions.
For the couplings of the light Higgs boson to third generation fermions we find
\begin{eqnarray} \label{eq:yt}
 \kappa_t &\equiv& \frac{Y_t}{Y_t^\text{SM}} = \frac{c_\alpha}{s_\beta} + \mathcal{O}\left(\frac{m_c}{m_t}\right)\times \frac{t_\beta}{s_\beta^2} c_{\beta-\alpha} ~, \\  \label{eq:yb}
 \kappa_b &\equiv& \frac{Y_b}{Y_b^\text{SM}} = \frac{c_\alpha}{s_\beta} + \mathcal{O}\left(\frac{m_s}{m_b}\right)\times \frac{t_\beta}{s_\beta^2} c_{\beta-\alpha} ~, \\  \label{eq:ytau}
 \kappa_\tau &\equiv& \frac{Y_\tau}{Y_\tau^\text{SM}} = \frac{c_\alpha}{s_\beta} + \mathcal{O}\left(\frac{m_\mu}{m_\tau}\right)\times \frac{t_\beta}{s_\beta^2} c_{\beta-\alpha} ~.
\end{eqnarray}
Note that the bulk of the correction with respect to the SM prediction is universal for the top, the bottom and the tau ($c_\alpha/s_\beta$), and are the same as in a 2HDM type I. The higher order terms are suppressed by small fermion mass ratios and can have order one CP violating phases. They can become relevant in the large $\tan\beta$ regime.

As in any other 2HDM, the reduced couplings of the light Higgs to the weak gauge bosons are given by
\begin{equation}
 \kappa_W = \kappa_Z \equiv \kappa_V = s_{\beta - \alpha} ~.
\end{equation}
The couplings of the Higgs to the lighter fermion generations are also modified. The expressions for the second generation read
\begin{eqnarray}
 \kappa_\mu &\equiv& \frac{Y_\mu}{Y_\mu^\text{SM}} = -\frac{s_\alpha}{c_\beta} + \mathcal{O}\left(\frac{m_\mu}{m_\tau}\right) \times \frac{t_\beta}{s_\beta^2} c_{\beta-\alpha} ~, \\
 \kappa_c &\equiv& \frac{Y_c}{Y_c^\text{SM}} = -\frac{s_\alpha}{c_\beta} + \mathcal{O}\left(\frac{m_c}{m_t}\right)  \times \frac{t_\beta}{s_\beta^2} c_{\beta-\alpha} ~, \\
 \kappa_s &\equiv& \frac{Y_s}{Y_s^\text{SM}} = -\frac{s_\alpha}{c_\beta} + \mathcal{O}\left(\frac{m_s}{m_b}\right) \times \frac{t_\beta}{s_\beta^2} c_{\beta-\alpha}  ~.
\end{eqnarray}
Analogous expressions hold for the first generation, with second generation masses replaced by first generation masses. 
The couplings to the first and second generation depend in a different way on $\alpha$ and $\beta$ as compared to the couplings of the third generation. This is a distinct feature of our framework in comparison to 2HDMs with natural flavor conservation~\cite{Glashow:1976nt} or flavor alignment~\cite{Pich:2009sp, Altmannshofer:2012ar}, which predict modifications of the couplings that are universal across the generations.
The corrections to the couplings for all first and second generation fermions are still universal, up to terms proportional to small ratios of fermion masses. Such terms are particularly small for the first generation.
Generically, all higher order terms can have order one CP violating phases.
Note that in the absence of mixing between the scalar components of the two Higgs doublets ($\alpha = 0$), the 125~GeV Higgs does not couple at all to the first and second generation. For large $\tan\beta$ and away from the decoupling or alignment limit $\cos(\alpha -\beta) = 0$, the couplings can deviate substantially from the SM prediction and can even be significantly enhanced. 

\begin{figure}[tb]
\centering
\includegraphics[width=0.45\textwidth]{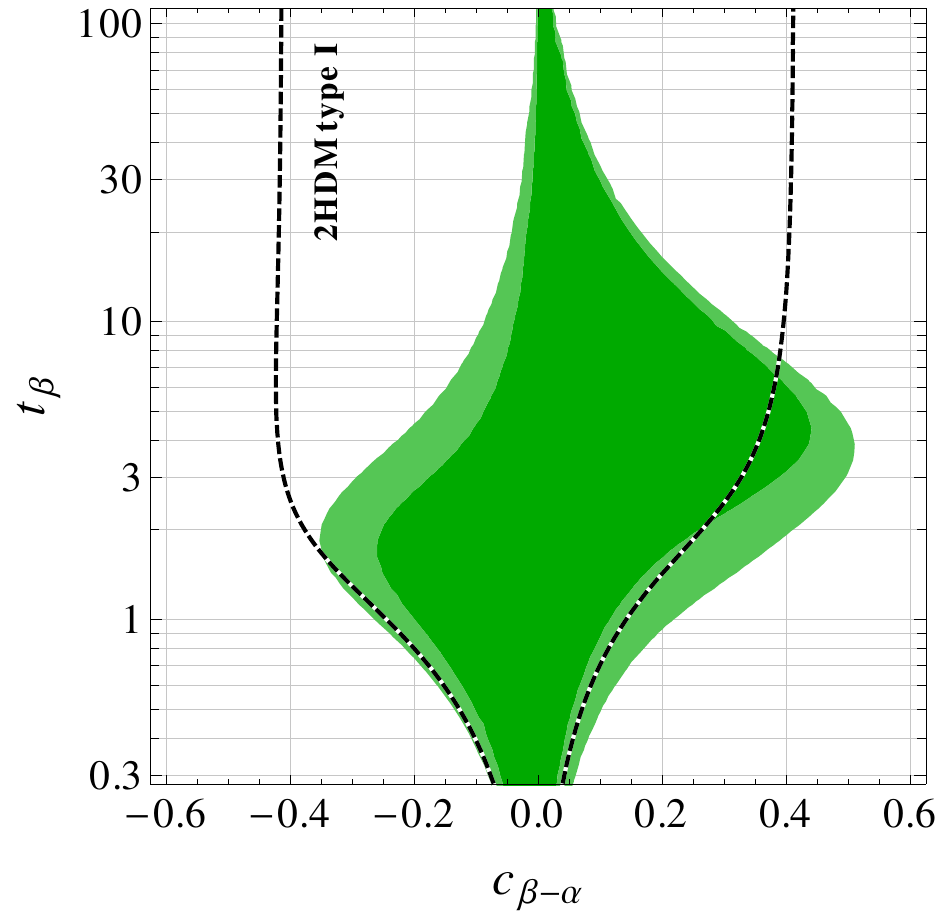}
\caption{Allowed region in the $\cos(\beta-\alpha)$ vs. $\tan\beta$ plane from measurements of the 125~GeV Higgs rates at the LHC. The dark green and light green regions correspond to the $1\sigma$ and $2\sigma$ allowed regions, allowing the $\mathcal{O}(m_\text{2nd}/m_\text{3rd})$ terms in the relevant Higgs couplings to float between $- 3 m_\text{2nd}/m_\text{3rd}$ and $+3 m_\text{2nd}/m_\text{3rd}$. The dashed line corresoponds to the $2\sigma$ contour in a 2HDM type I.}
\label{fig:higgs_couplings}
\end{figure}

Measurements of Higgs production and decay rates can be used to constrain the allowed ranges for the angles $\alpha$ and $\beta$. We use the results for the Higgs signal strengths given in~\cite{Khachatryan:2016vau} to construct a $\chi^2$ function depending on the couplings of the Higgs to vector bosons, top, bottom and charm quarks, as well as taus and muons, including the given correlations of the signal strength uncertainties. The results in~\cite{Khachatryan:2016vau} consist of 20 combinations of five production mechanisms (gluon fusion, vector boson fusion, producution in association with $W$, $Z$ and $t \bar t$), and five branching ratios ($WW$, $ZZ$, $\gamma\gamma$, $\tau^+\tau^-$, $b\bar b$) that combine ATLAS and CMS measurements at 7 and 8~TeV. To construct the signal strengths in our model, we use the SM production cross sections and branching ratios for a 125~GeV Higgs from~\cite{Heinemeyer:2013tqa} and reweight them with the appropriate combinations of coupling modifiers. We add to the $\chi^2$ 
also 
the 
13 TeV bound on the signal strength into muons~\cite{ATLAS:2016zzs} using the modified inclusive Higgs production cross section at 13~TeV, assuming vanishing correlation with the signal strength measurements from~\cite{Khachatryan:2016vau}.

The derived constraint in the $\cos(\beta-\alpha)$ vs. $\tan\beta$ plane is shown in Fig.~\ref{fig:higgs_couplings}. 
The dark (light) green region correspond to $\Delta \chi^2 = \chi^2 - \chi^2_\text{SM} < 1 (4)$, allowing the $\mathcal{O}(m_\text{2nd}/m_\text{3rd})$ terms in the involved Higgs couplings to float between $- 3 m_\text{2nd}/m_\text{3rd}$ and $+3 m_\text{2nd}/m_\text{3rd}$.
If we set the mass-suppressed corrections to the third generation couplings to zero and we completely neglect the modifications of the charm and muon coupling, the constraint in the $\cos(\beta-\alpha)$ vs. $\tan\beta$ plane coincides with the constraints in a 2HDM type I. The corresponding $\Delta \chi^2 = 4$ contour is shown with a dashed line and qualitatively reproduces the 2HDM type I constraints given in the ATLAS and CMS analyses~\cite{Aad:2015pla, CMS:2016qbe}.

We find that the modifications of the charm and muon couplings have an important impact on the fit.
For large $\tan\beta$ and away from the decoupling or alignment limit, $\cos(\beta-\alpha)= 0$, the charm and muon couplings can be strongly enhanced, leading to a substantially larger total width of the Higgs and a largely enhanced branching ratio into muons. 
For moderate and large values of $\tan\beta$, the allowed region therefore differs significantly from the 2HDM type I case.
In the remaining parts of this paper we take into account the constraints coming from the measurments of the 125~GeV Higgs rates by imposing $\Delta \chi^2 < 4$.

\bigskip
In addition to the modified SM couplings of the light Higgs, our framework also gives rise to the flavor-violating couplings in Eq.~(\ref{eq:yllp}).  
The corresponding flavor-violating decays of the light Higgs boson have branching ratios of\footnote{Throughout the paper, we will denote the flavor-changing decays $\bar ff^\prime+f\bar f^\prime$, simply as $ff^\prime$.}
\begin{eqnarray}
 \text{BR}(h \to ff^\prime) &=& \text{BR}(h \to f\bar f^\prime) + \text{BR}(h \to \bar f f^\prime) \nonumber \\
 &=& \frac{m_h}{8 \pi \Gamma_h} \Big( |Y_{ff^\prime}|^2 + |Y_{f^\prime f}|^2 \Big) ~,
\end{eqnarray}
where $\Gamma_h$ is the total Higgs width and we have neglected tiny phase space effects.

For $h \to \tau\mu$ and $h \to \tau e$ this gives generically branching ratios of the order of
\begin{eqnarray}
 \text{BR}(h \to \tau\mu) &\sim& \text{BR}(h \to \mu^+\mu^-) \sim \frac{m_\mu^2}{3 m_b^2} \sim 10^{-3}, \\ 
 \text{BR}(h \to \tau e) &\sim& \frac{m_e^2}{m_\mu^2} \times \text{BR}(h \to \tau\mu) \sim 10^{-7} ~.
\end{eqnarray}
This implies that $h \to \tau \mu$ can be at an experimentally accessible level and the model could even explain the observed excess in $h \to \tau \mu$ searches at CMS~\cite{Khachatryan:2015kon}. The decay $h \to \tau e$, on the other hand, is generically well below the foreseeable experimental sensitivities.
The prediction for $h \to \mu e$ is even smaller
\begin{equation}
 \text{BR}(h \to \mu e) \sim \frac{m_e^2}{m_\tau^2} \times \text{BR}(h \to \tau\mu) \sim 10^{-10}~.
\end{equation}

In the quark sector the $h \to bs$ mode has generically the largest branching ratio
\begin{equation}
 \text{BR}(h \to bs) \sim |V_{cb}|^2 \times \text{BR}(h \to b \bar b) \sim 10^{-3}~.
\end{equation}
In view of the large $h \to b \bar b$ background, this is too small to be seen at the LHC. Other flavor-changing Higgs decays into quarks are even smaller and even more challenging to detect.

\section{Heavy Neutral Higgs Production and Decays} \label{sec:neutralhiggs}

As we saw at the end of Sec.~\ref{sec:rank1}, several of the heavy Higgs couplings depend significantly on the entries of the $m^\prime$ mass matrices, which are free parameters.
To simplify our discussion of the heavy Higgs phenomenology we chose a constrained setup with a reduced set of free parameters.

In the $\lambda^\prime$ Yukawa couplings for the leptons and up-type quarks (see Eq.~(\ref{eq:l_texture})), we set
\begin{eqnarray} \label{eq:texture_restricted_1}
 \lambda^{\prime \ell,u}_{11} = \lambda^{\prime \ell,u}_{12} = \lambda^{\prime \ell,u}_{13} = \lambda^{\prime \ell,u}_{21} = \lambda^{\prime \ell,u}_{31} \,, \\
 \lambda^{\prime \ell,u}_{22} = \lambda^{\prime \ell,u}_{23} = \lambda^{\prime \ell,u}_{32} = \lambda^{\prime \ell,u}_{33} \,. \label{eq:texture_restricted_2}
\end{eqnarray}
For any given $\tan\beta$, the values of these parameters are fixed such to reproduce the observed electron, muon, up, and charm masses (we use $\overline{\text{MS}}$ masses at a scale $\mu = 500$~GeV).

For the $\lambda^\prime$ Yukawa couplings for the down-type quarks, we use the texture in Eq.~(\ref{eq:d_texture}) with
\begin{eqnarray}
 \lambda^{\prime d}_{11} = \lambda^{\prime d}_{21} = \lambda^{\prime d}_{31} \,, \\
 \lambda^{\prime d}_{22} = \lambda^{\prime d}_{32} = \lambda^{\prime d}_{33} \,. \label{eq:texture_restricted_3}
\end{eqnarray}
The down and strange masses, together with the CKM angles fix all entries of the $\lambda^{\prime d}$ matrix for a given value of $\tan\beta$. 
With the above assumptions, the Higgs mixing angle $\alpha$ and $\tan\beta$ completely determine all Higgs couplings.

All results we will present in the following depend very little on the choice of the $\lambda^\prime_{1i}$ and $\lambda^\prime_{i1}$ parameters. However, some results do depend on the chosen values in the $2-3$ block of the $\lambda^\prime$ Yukawa couplings.    
Whenever this dependence is strong, we will comment on the impact a perturbation would have around the described restricted setup.

\subsection{Branching Ratios}

\begin{figure*}[tb]
\centering
\includegraphics[width=0.45\textwidth]{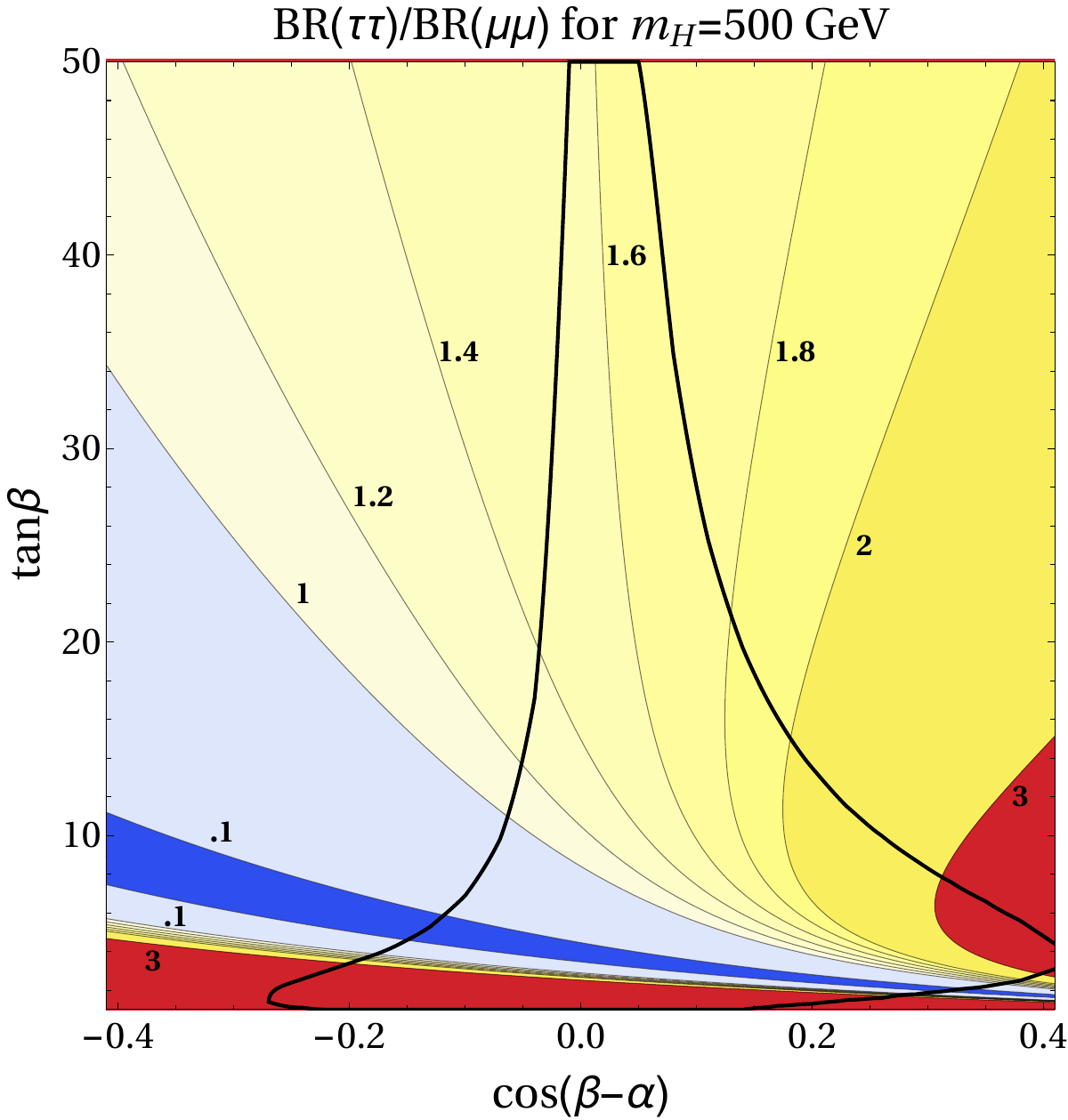} ~~~
\includegraphics[width=0.45\textwidth]{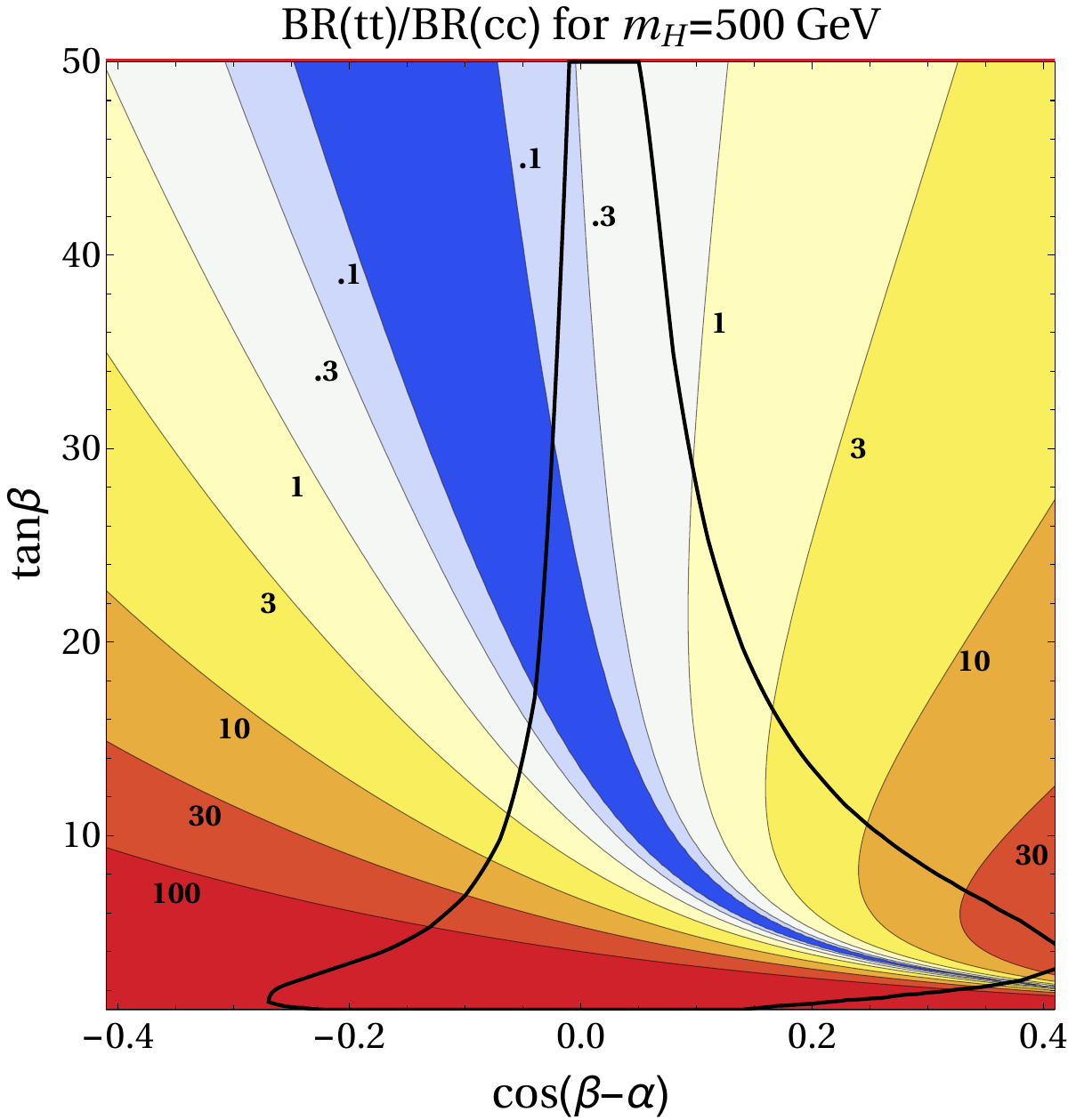}
\caption{Ratio of branching ratios $H \to \tau^+ \tau^-$ over $H \to \mu^+\mu^-$ (left)
and $H \to t \bar t$ over $H \to c \bar c$ (right) in the $\tan\beta$ vs. $\cos(\beta-\alpha)$ plane for a heavy Higgs with mass $m_H = 500$~GeV. Outside the black solid contours, the 125~GeV Higgs rates are in conflict with LHC data.}
\label{fig:ratios}
\end{figure*}

In addition to well-studied heavy Higgs decays $H \to WW/ZZ$, $A \to Zh$, and $A/H \to t \bar t, b \bar b, \tau^+\tau^-$, we are particularly interested in decays involving lighter fermion flavors like $A/H \to c \bar c, \mu^+\mu^-$ and the flavor-violating decays $A/H \to tc, \tau\mu$.
We assume that the heavy Higgs sector is approximately degenerate, $m_H \simeq m_A \simeq m_{H^\pm}$, such that no two body decay modes involving heavy Higgses in the final state are kinematically allowed. 
We also assume that the triple Higgs couplings $Hhh$ and $Ahh$ are sufficiently small such that we can neglect the $H \to hh$ and $A \to hh$ decay modes.\footnote{The $A \to hh$ decay is automatically zero in the absence of CP violation in the Higgs sector, while, in the almost decoupling or alignment limit and at large values of $\tan\beta$, $H \to hh$ depends mainly on the $\lambda_7$ quartic coupling that is equal to zero if the two doublets have an opposite $Z_2$ charge (see e.g. \cite{Branco:2011iw} for the definition of $\lambda_7$).} 
For the calculation of the Higgs branching ratios we use leading-order expressions for all relevant partial widths.

The characteristic flavor structure of the model can be easily grasped by looking at ratios of branching ratios involving second and third generation fermions.
For example, in 2HDMs with natural flavor conservation or flavor alignment one finds
\begin{eqnarray}
\frac{\text{BR}(A \to \tau^+\tau^-)}{\text{BR}(A \to \mu^+\mu^-)} = \frac{\text{BR}(H \to \tau^+\tau^-)}{\text{BR}(H \to \mu^+\mu^-)} = \frac{m_\tau^2}{m_\mu^2} \simeq 300 \,, \\
\frac{\text{BR}(A \to t\bar t)}{\text{BR}(A \to c \bar c)} \simeq \frac{\text{BR}(H \to t \bar t)}{\text{BR}(H \to c \bar c)} \simeq \frac{m_t^2}{m_c^2} \simeq 7 \times 10^4 \,,
\end{eqnarray}
where, for illustration, we used running $\overline{\text{MS}}$ quark masses at the scale $\mu = 500$~GeV and neglected phase space effects that might be relevant in the decay to top quarks.  
In our setup, the above relations can be strongly violated. For the pseudoscalar $A$ we obtain
\begin{eqnarray} \label{eq:Rtaumu}
 \frac{\text{BR}(A \to \tau^+\tau^-)}{\text{BR}(A \to \mu^+\mu^-)} &\simeq& \frac{m_\tau^2}{m_\mu^2} \frac{1}{t_\beta^4} \left( 1 - \frac{t_\beta}{s_\beta c_\beta} \frac{m_{\tau\tau}^\prime}{m_\tau}  \right)^2  \,, \\ \label{eq:Rtc}
 \frac{\text{BR}(A \to t \bar t)}{\text{BR}(A \to c \bar c)} &\simeq& \frac{m_t^2}{m_c^2} \frac{1}{t_\beta^4} \left( 1 - \frac{t_\beta}{s_\beta c_\beta} \frac{m_{tt}^\prime}{m_t}  \right)^2  \,,
\end{eqnarray}
where we neglected respectively $\mathcal{O}(m_\mu/m_\tau)$ and $\mathcal{O}(m_c/m_t)$ corrections. The expressions~(\ref{eq:Rtaumu}) and~(\ref{eq:Rtc}) also hold for the heavy scalar $H$, up to corrections of $\mathcal{O}(v_W^2/m_A^2)$.
For moderate $t_\beta$ we can neglect the terms proportional to $m_{\tau\tau}^\prime$ and $m_{tt}^\prime$ and obtain the ratios $m_\tau^2/(m_\mu^2 t_\beta^4)$ and $m_t^2/(m_c^2 t_\beta^4)$, respectively. For large $t_\beta$, the terms proportional to $m_{\tau\tau}^\prime$ and $m_{tt}^\prime$ are dominant and we find the ratios $(m_{\tau\tau}^\prime)^2/m_\mu^2$ and $(m_{tt}^\prime)^2/m_c^2$. In all cases, the ratios of branching ratios can be of $\mathcal{O}(1)$.

\begin{figure*}[tb]
\centering
\includegraphics[width=0.45\textwidth]{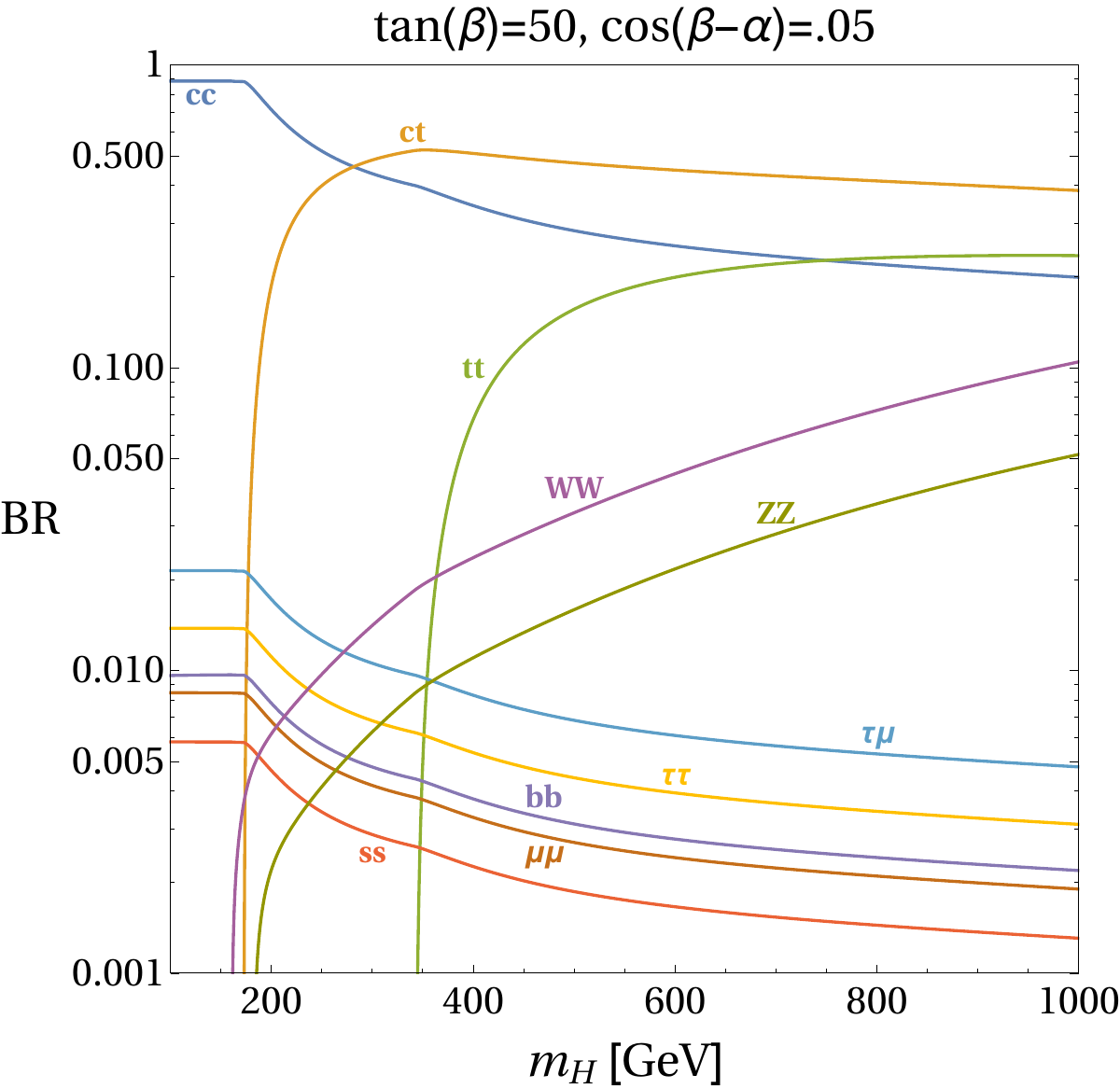} ~~~
\includegraphics[width=0.45\textwidth]{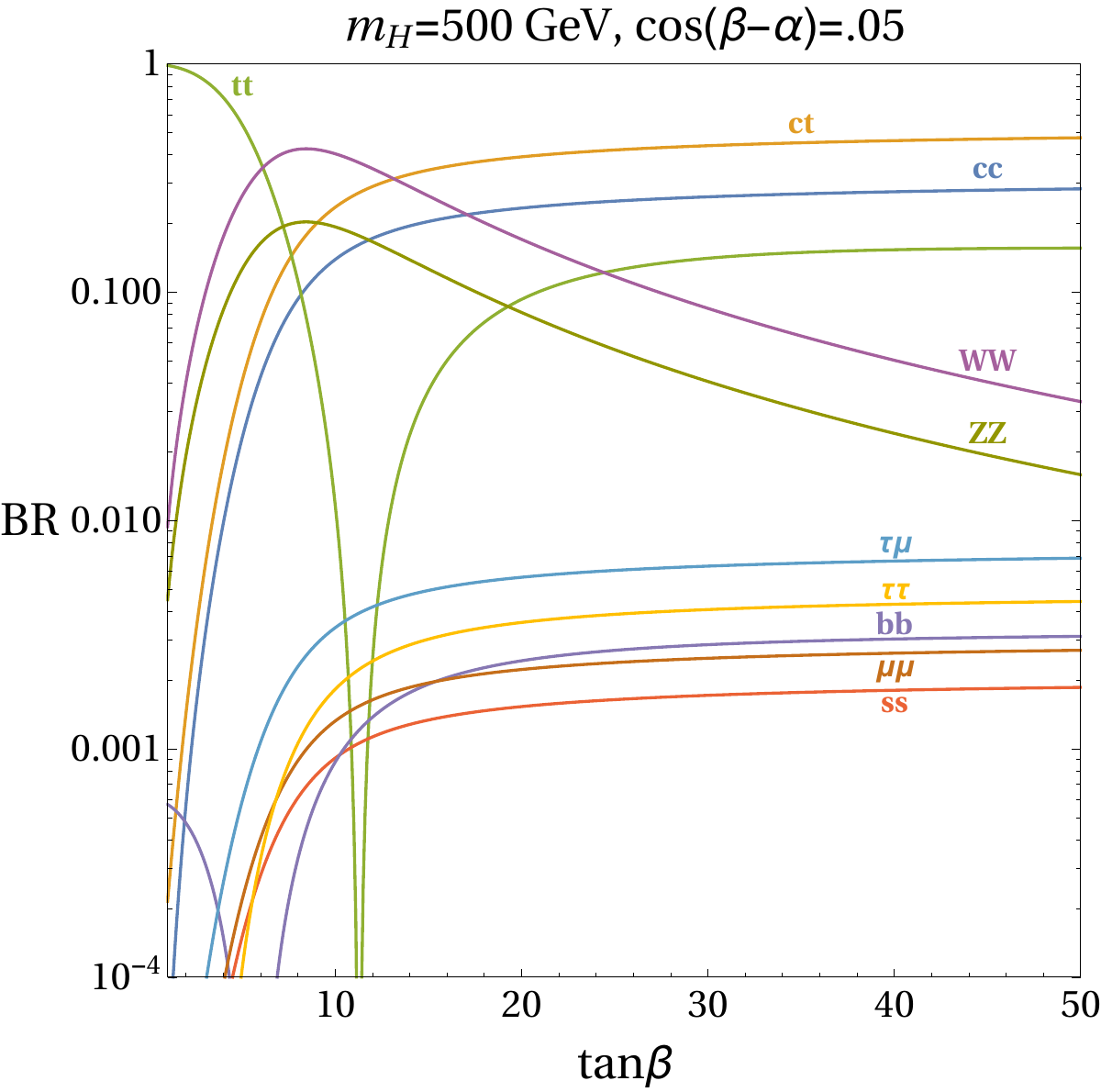}
\caption{Branching ratios of the scalar $H$ as a function of its mass $m_{H}$ for fixed $\tan\beta = 50$ (left) and as a function of $\tan\beta$ for fixed Higgs mass $m_H = 500$~GeV (right). In both plots we set $\cos(\beta-\alpha) = 0.05$.}
\label{fig:BR}
\end{figure*}

This is illustrated in Fig.~\ref{fig:ratios}, that shows the ratio of $\tau^+\tau^-$ and $\mu^+\mu^-$ branching ratios (left) as well as of $t \bar t$ and $c \bar c$ branching ratios (right) of the scalar $H$ in the plane of $\cos(\beta-\alpha)$ vs. $\tan\beta$ for a scalar mass of $m_H = 500$~GeV. The values of the pseudoscalar branching ratios can be obtained from the figure, by fixing $\cos(\beta-\alpha)=0$.
Outside the black solid contours, the 125~GeV Higgs rates are in conflict with LHC data (see Fig.~\ref{fig:higgs_couplings}). 

Note that for small and moderate $\tan\beta$, these ratios of branching ratios are not very sensitive to our choice of Yukawa matrices in Eqs.~(\ref{eq:texture_restricted_1}) and~(\ref{eq:texture_restricted_2}). For large $\tan\beta$, however, they are determined by $m^\prime_{tt}$ and $m^\prime_{\tau\tau}$ which are in general free parameters. The values shown in Fig.~\ref{fig:ratios} in the large $\tan\beta$ regime should therefore be regarded as typical expectations that could be larger or smaller by a factor of few. Overall, we see that the ratios are much smaller than in models with natural flavor conservation, minimal flavor violation or flavor alignment.

Similarly, also the flavor-violating decays into the $\tau \mu$ and $t c$ final states can have sizable branching ratios. For the pseudoscalar $A$ we have approximately
\begin{eqnarray} \label{fv1}
 \frac{\text{BR}(A \to \tau\mu)}{\text{BR}(A \to \mu^+\mu^-)} &\simeq& \frac{1}{s_\beta^4} \frac{(m_{\mu\tau}^\prime)^2 + (m_{\tau\mu}^\prime)^2}{m_\mu^2}  \,, \\\label{fv2}
 \frac{\text{BR}(A \to tc)}{\text{BR}(A \to c \bar c)} &\simeq& \frac{1}{s_\beta^4} \frac{(m_{ct}^\prime)^2 + (m_{tc}^\prime)^2}{m_c^2}  \,,
\end{eqnarray}
and similar expressions hold for the scalar $H$. The $m^\prime$ entries which determine \eqref{fv1} and \eqref{fv2} are in general free parameters. Typically, we expect the flavor-violating branching ratios to be within a factor of few of the flavor-diagonal decays $\mu^+\mu^-$ and $c \bar c$, respectively.

The plots in Fig.~\ref{fig:BR} show the branching ratios of the scalar $H$ as a function of $m_H$ for fixed $\tan\beta = 50$ (left) and as a function of $\tan\beta$ for fixed $m_H = 500$~GeV (right). In both plots we set $\cos(\beta-\alpha) = 0.05$ to satisfy constraints from the 125~GeV Higgs coupling measurements as discussed in Sec.~\ref{sec:SMhiggs}. 
For low values of $\tan\beta$, the decay into the $t \bar t$ final state dominates if kinematically allowed.
At large $\tan\beta$, decays into $t \bar t$, $c \bar c$ and the flavor-violating mode $t c$ have the largest branching ratios. Typically, these decay modes have branching ratios within a factor of few from each other. The sudden and strong suppression of the $t \bar t$ branching ratio is due to an accidental cancellation between the two terms entering the coupling of the heavy scalar to tops (cf. Eq. (\ref{exlept2}) and text below). The coupling $Y^H_{tt}$  vanishes at approximately $\tan\beta \simeq 11$. The value of $\tan\beta$ where such a cancellation occurs can shift by a factor of few, depending on the $m^\prime_{tt}$ parameter. For the opposite sign of $m^\prime_{tt}$, the cancellation does not occur instead. A similar, but less prominent, phenomenon happens for the $b\bar b$ branching ratio: for our choices of parameters, the coupling $Y^H_{bb}$ vanishes at $\tan\beta \simeq 5.6$. 

For $\cos(\beta-\alpha) = 0.05$, the decay into $WW$ and $ZZ$ can be non-negligible. Typically we find branching ratios of the order of few-10s \%. For moderate $\tan\beta$, these decays can even dominate. 
Concerning the leptonic decay modes $\tau^+\tau^-$, $\mu^+\mu^-$, and $\tau \mu$, for moderate and large values of $\tan\beta$ we find typical branching ratios at the level of $10^{-3}$ to $10^{-2}$. Branching ratios involving second and third generation down-type quarks (only $b \bar b$ and $s \bar s$ are shown in the plots) are generically at a comparable level. 
For moderate and large $\tan\beta$, the values of the flavor-violating partial widths and the partial width to $t \bar t$, $b \bar b$, and $\tau^+\tau^-$ depend on the $m^\prime$ mass matrices. Therefore,
perturbing the $m^\prime$ matrices around the ansatz based on Eqs.~(\ref{eq:texture_restricted_1}) -~(\ref{eq:texture_restricted_3}), can increase or suppress the various $H$ branching ratios by a factor of few.

The branching ratios of the pseudoscalar $A$ show qualitatively a very similar behavior and, for this reason, we do not show the corresponding figures.
For a mass of $A$ above the $t \bar t$ threshold, the $t \bar t$ branching ratio is dominant for low $\tan \beta$, while for large $\tan\beta$ also the decay to the $c \bar c$ final state and the flavor-violating decay to $tc$ become comparable in size. Decays involving second and third generation leptons are all comparable and typically at a level of few~$\times\ 10^{-3}$. In contrast to the heavy scalar, $H$, the heavy pseudoscalar, $A$, cannot decay at tree level into a pair of gauge bosons. Instead, the decay $A \to Z h$ is possible. The corresponding partial decay width is proportional to $\cos^2(\beta-\alpha)$. For $\cos(\beta-\alpha) = 0.05$, the $A \to Z h$ branching ratio is around a few \%. For heavy pseudoscalar masses and moderate values of $\tan\beta$, this decay mode might dominate.

\subsection{Production Cross Sections}

\begin{figure*}[tb]
\centering
\includegraphics[width=0.28\textwidth]{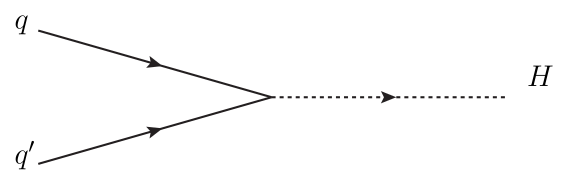}~~~
\includegraphics[width=0.36\textwidth]{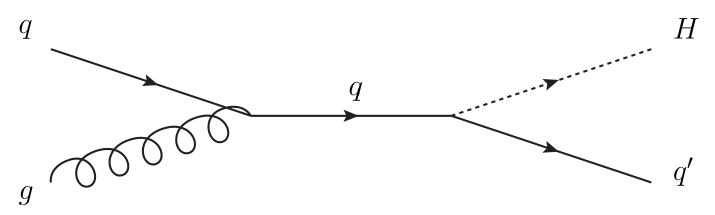}~~~
\includegraphics[width=0.28\textwidth]{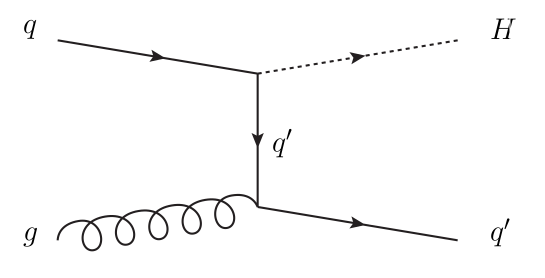}
\caption{Feynman diagrams for the most interesting (and novel) production modes of the heavy neutral Higgs bosons. Left: production from quark quark fusion (mainly coming from $c \bar c$); Center and Right: production in association with a top/bottom with the main contributions coming from flavor-changing diagrams where the initial state $q$ is a charm/strange quark.}
\label{fig:diagrams}
\end{figure*}

\begin{figure*}[tb]
\centering
\includegraphics[width=0.45\textwidth]{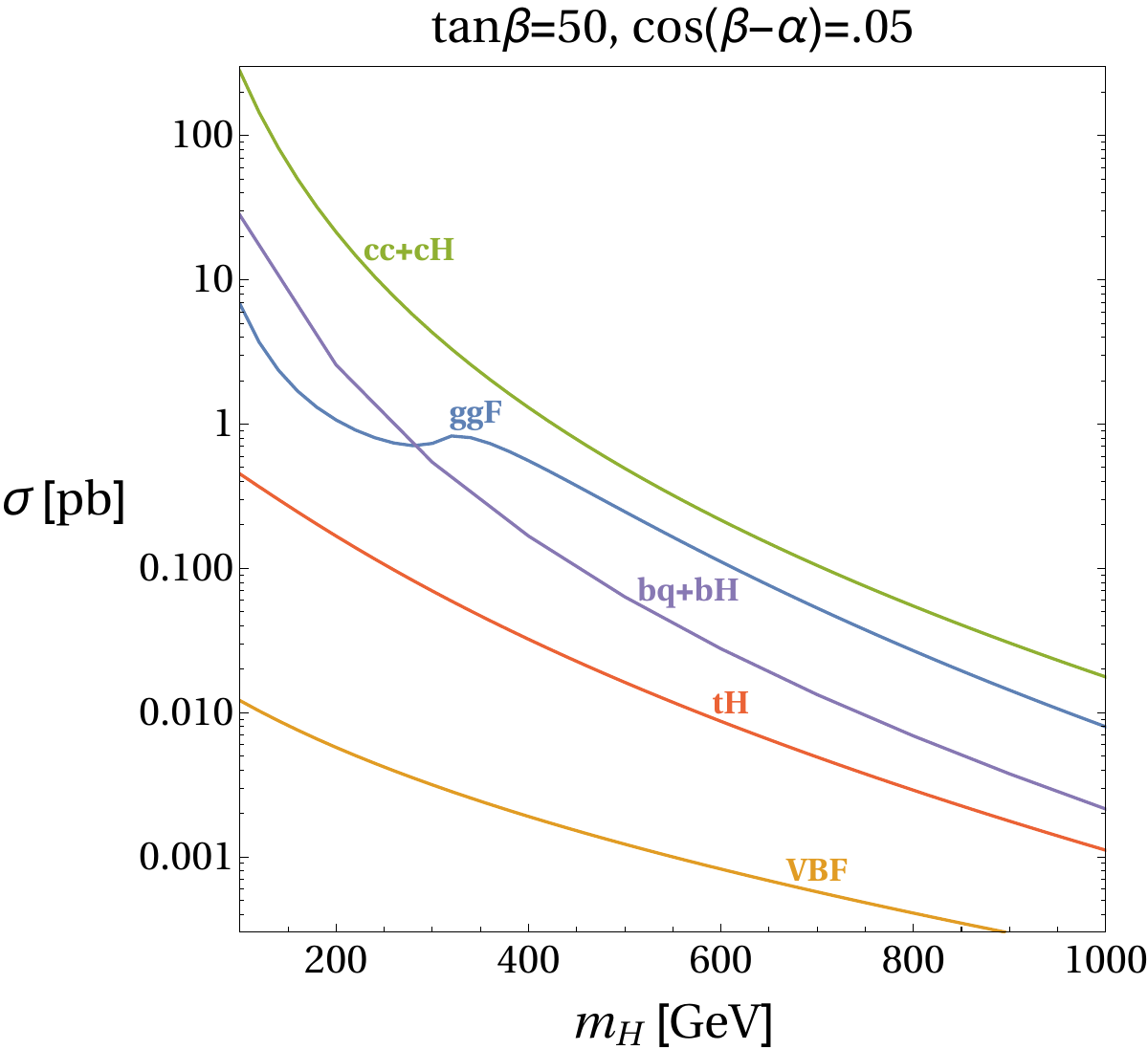} ~~~
\includegraphics[width=0.45\textwidth]{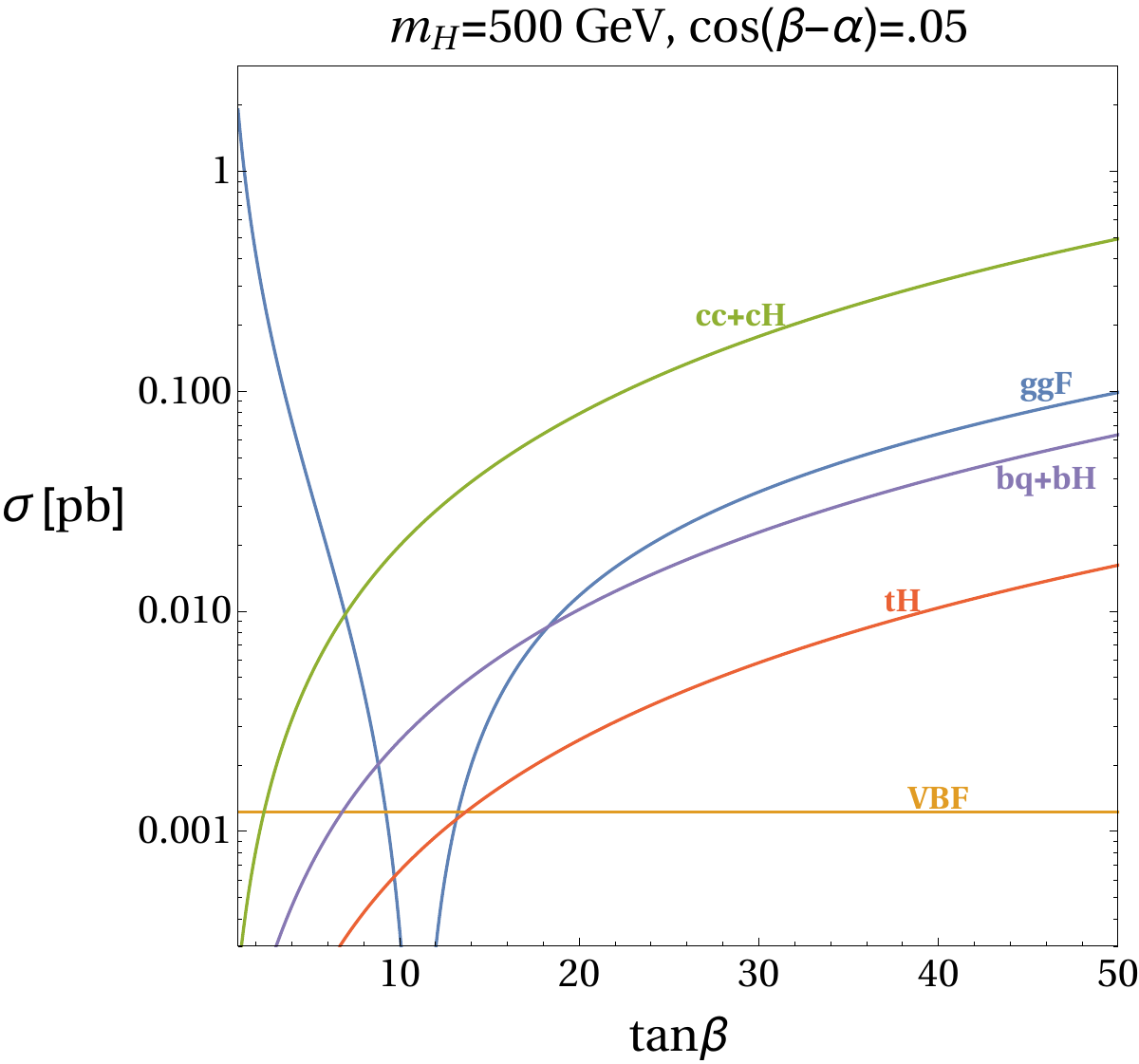}
\caption{Production cross sections of the scalar $H$ at 13~TeV proton~proton collisions as a function of the scalar mass $m_H$ for fixed $\tan\beta = 50$ (left) and as a function of $\tan\beta$ for fixed scalar mass $m_H = 500$~GeV (right).  The $cc+cH$ curves include both the $c\bar c$ and the associated $cH$ and $\bar c H$ production cross sections. In both plots we set $\cos(\beta-\alpha) = 0.05$.}
\label{fig:sigma}
\end{figure*}

We consider various production mechanisms of the heavy neutral Higgs bosons, including gluon fusion, vector boson fusion, production in association with weak vector bosons and light quarks, production from a $c \bar c$ initial state, and also flavor-violating production in association with a top or a bottom quark. Example diagrams for the novel processes are shown in Fig.~\ref{fig:diagrams}.

Throughout all regions of parameter space, we find that the gluon fusion production cross section is dominated by the top quark loop.
The bottom quark loop gives a \% level correction, which is included in the numerics. Also the charm quark loop gives generically only a small correction (approximately 5\% in the large $\tan\beta$ regime for a Higgs mass of 500~GeV).
In our numerical analysis we use leading-order expressions for the gluon fusion production cross sections that we convolute with MMHT2014 NNLO PDFs~\cite{Harland-Lang:2014zoa}. We set the renormalization and factorization scales to 500~GeV and multiply the cross section with a constant $K$ factor of 2.5 to approximate higher order corrections.

The vector boson fusion production cross section of the heavy scalar $H$ is suppressed by $\cos^2(\beta-\alpha)$, if compared to the corresponding SM Higgs cross section. In the regions of parameter space that are compatible with the observed 125~GeV Higgs rates, vector boson fusion is therefore typically subdominant. The same applies to production of $H$ in association with weak gauge bosons.
In our numerical analysis we use the corresponding production cross sections given in~\cite{Heinemeyer:2013tqa} rescaled by the appropriate factor $\cos^2(\beta-\alpha)$. 
The pseuedoscalar $A$ does not couple to weak gauge bosons and thus cannot be produced in vector boson fusion or in association with $W$ or $Z$ bosons. 
It can be produced in association with the light Higgs: $q \bar q \to Z^* \to A h$. The corresponding cross section is proportional to $\cos^2(\beta-\alpha)$ and therefore small. 

Due to the enhanced couplings of the heavy Higgses to second generation quarks, we expect sizable production of $H$ and $A$ from a $c \bar c$ initial state. Production in association with a $c$ or $\bar c$ from a gluon+charm initial state is also sizable. In such a case the associated charm might escape detection giving rise to collinear logarithms which need a careful analysis. To this end we follow~\cite{Harlander:2003ai} and do not consider production in association with a $c$ or a $\bar{c}$ as a separate production channel but as a NLO correction to $c\bar{c}$. For our calculations we use the corresponding parton level expressions in~\cite{Harlander:2003ai} up to NLO accuracy and convolute them with MMHT2014 NNLO PDFs.

We also consider production of the heavy scalar and pseudoscalar in association with with a top quark and with a bottom quark. These processes are mainly initiated by flavor-violating $tc$ and $bs$ couplings, respectively (see central and right panel of Fig. \ref{fig:diagrams}). For the production in association with a top quark we use LO expressions for the parton level cross sections and MMHT2014 NNLO PDFs. For the production in association with a bottom quark we instead perform a LO computation, using MadGraph5 \cite{Alwall:2014hca}. 

In Fig.~\ref{fig:sigma} we show the production cross sections of the scalar $H$ at 13~TeV proton-proton collisions as a function of $m_H$ for fixed $\tan\beta = 50$ (left) and as a function of $\tan\beta$ for fixed $m_H = 500$~GeV (right). In both plots we set $\cos(\beta-\alpha) = 0.05$. 
For a heavy Higgs mass of $m_H = 500$~GeV the inclusive production cross section can be few~$\times\ 100$~fb over a broad range of $\tan\beta$. The most important production modes are gluon fusion (denoted with ggF in the plots) and from processes where the Higgs couples to charm quarks $c\bar c \to H$, $g c \to H c$, and $g \bar c \to H \bar c$ (the sum of these modes is denoted with $cc+cH$ in the plots). Gluon fusion is dominant for small $\tan\beta$, while charm initiated production can dominate over the gluon fusion cross section for moderate and large values of $\tan\beta$. 
The strong suppression of the gluon fusion cross section for $\tan\beta \simeq 11$ is due to the same accidental cancellation in $Y^H_{tt}$ which leads to the suppression of BR($H\to t\bar t$) at this value of $\tan\beta$ (see discussion in the previous subsection). 

We find that production from $s \bar s$ (not shown in the plots) is suppressed by almost 2 orders of magnitude compared to $c \bar c$. The larger strange quark PDF cannot compensate for the much smaller coupling to the heavy Higgs proportional to $m_s$ vs. $m_c$.
For $\cos(\beta-\alpha) = 0.05$, production in vector-boson fusion is very small, with production cross sections ranging from 5.7~fb at a mass of $m_H = 200$~GeV to 0.22~fb at a mass of $m_H = 1$~TeV. Production in association with $W$ or $Z$ (not shown) is even smaller.  

The production of the heavy scalar in association with a bottom or a top can have appreciable cross sections at the level of 10s of fb for $m_H=500$ GeV, over a broad range of $\tan\beta$. In the bottom initiated production we include $bg\to Hb$, $\bar{b}g\to H\bar{b}$, $b\bar{b}\to H$, $b\bar{s}\to H$, $s\bar{b}\to H$, $sg\to Hb$, and $\bar{s}g\to H\bar{b}$, the latter two processes being the dominant ones, thanks to the strange quark PDF enhancement. 
For this reason in Fig.~\ref{fig:sigma} we label the bottom associated production by $bq+bH$. Overall, the cross section for the bottom associated production is, however, typically smaller than the one predicted in a 2HDM of type II. The $Ht$ associated cross section depends strongly on the free $m^\prime$ parameters and can easily be increased or decreased by a factor of few. 

The production modes of the pseudoscalar $A$ show a very similar behavior. Gluon fusion dominates for low $\tan\beta$, charm initiated production dominates for moderate and large values of $\tan\beta$. 
Production in association with top and bottom can have non-negligible cross sections. Vector boson fusion and production in association with vector bosons is absent for the pseudoscalar.

\section{Charged Higgs Production and Decays} \label{sec:chargedhiggs}

\subsection{Branching Ratios}

\begin{figure*}[tb]
\centering
\includegraphics[width=0.45\textwidth]{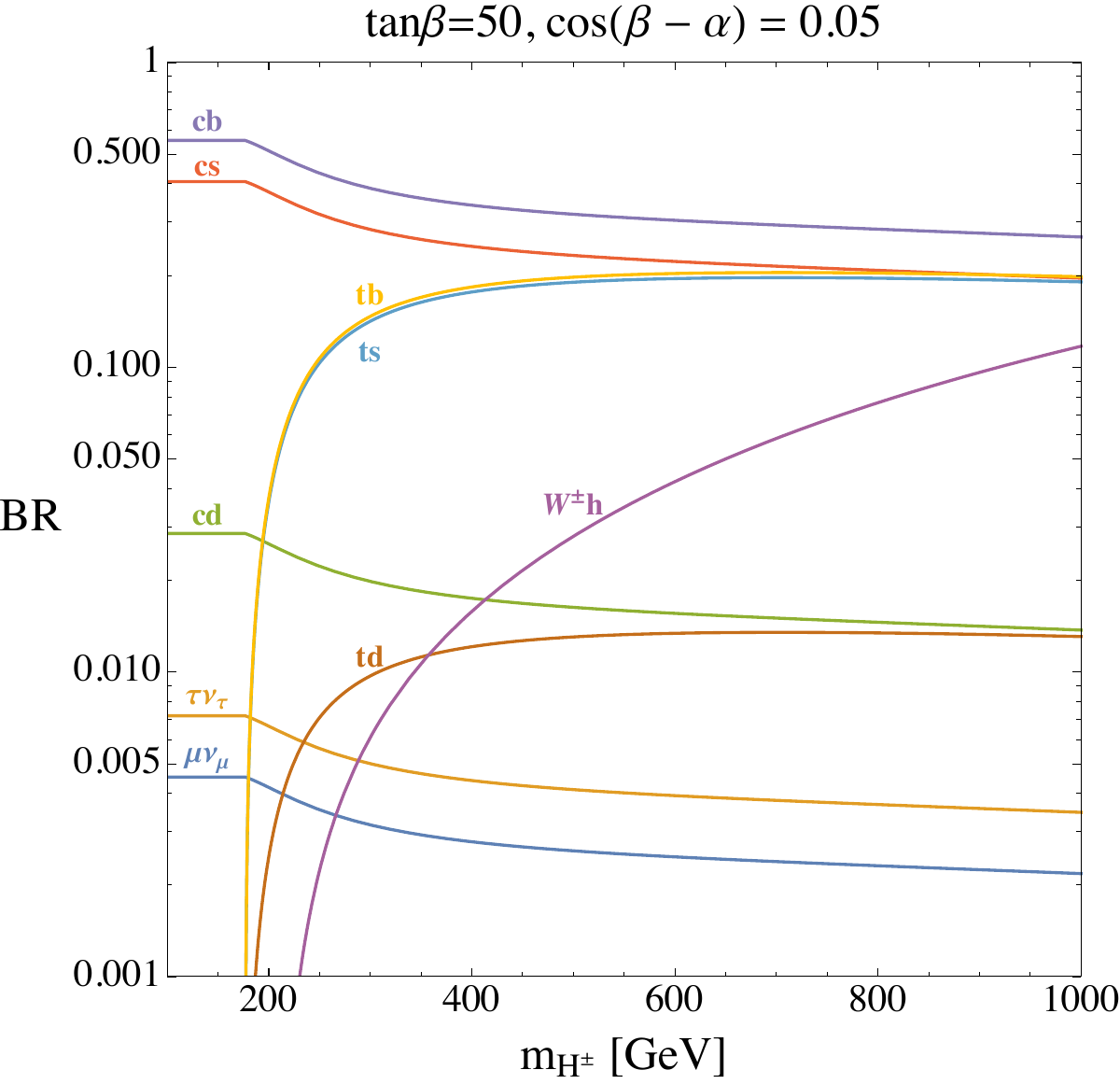} ~~~
\includegraphics[width=0.45\textwidth]{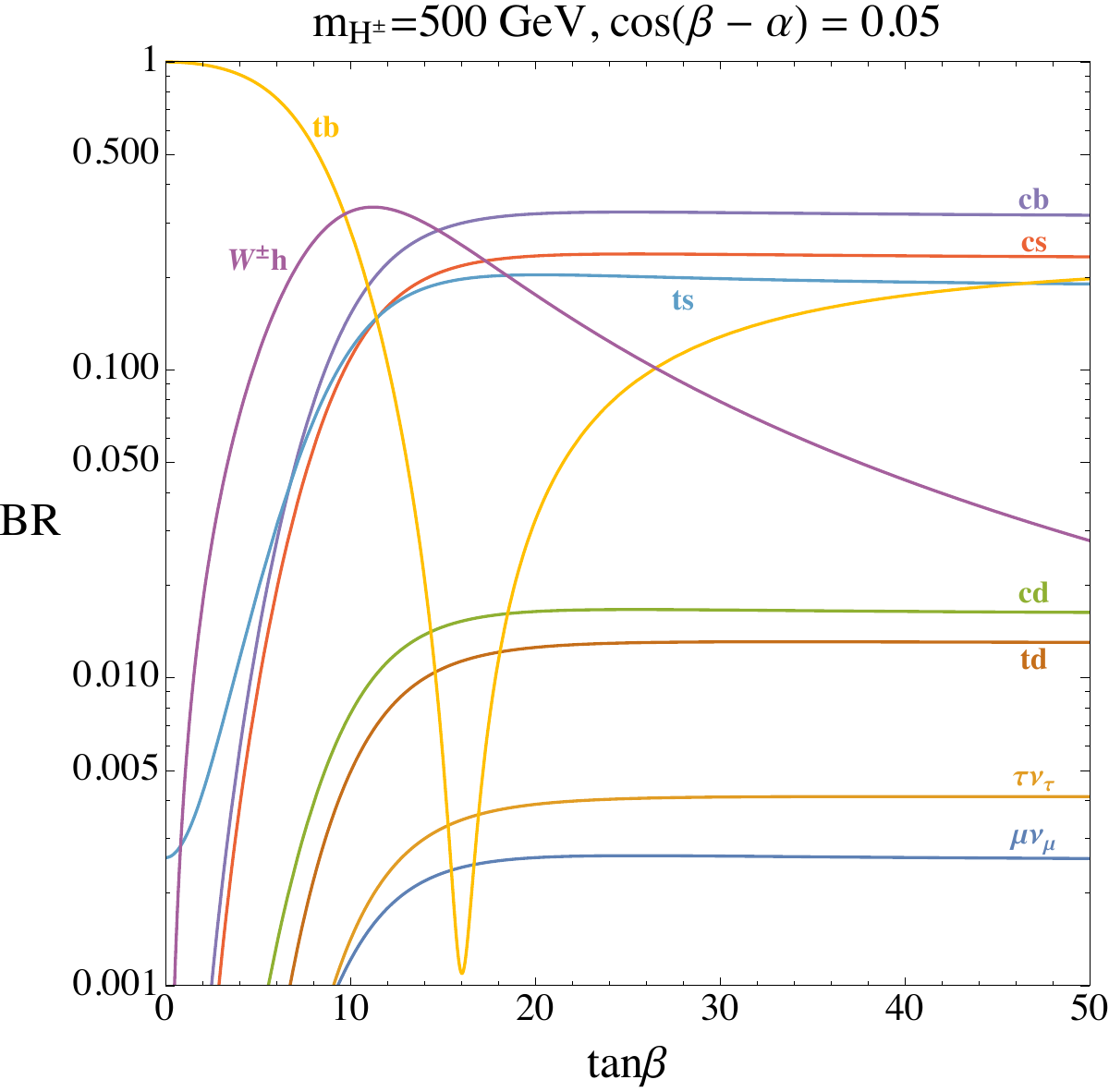}
\caption{Branching ratios of the charged Higgs $H^\pm$ as a function of the charged Higgs mass $m_{H^\pm}$ for fixed $\tan\beta = 50$ (left) and as a function of $\tan\beta$ for fixed Higgs mass $m_{H^\pm} = 500$~GeV (right). For both panels, we fix $\cos(\beta-\alpha)=0.05$.}
\label{fig:BRcharged}
\end{figure*}

Similarly to the neutral scalars, in addition to the well-studied $t b$ and $\tau\nu$ charged Higgs decay modes, we are interested in the flavor-violating decays, $c b$ and $t s$, as well as in the decays to second generations, $c s$ and $\mu \nu_\mu$. Particularly, from the charged Higgs couplings in (\ref{eq:chargedL}) - (\ref{eq:chargedLept}), we learn that the decay modes $t b$, $t s$, $c b$ and $c s$ should be of the same order, as long as they are kinematically open. The same observation holds also for the $\tau\nu_\tau$ and $\mu\nu_\mu$ decay modes, as opposed to the relation BR$(H^\pm \to \tau\nu_\tau)/{\rm{BR}}(H^\pm \to \mu\nu_\mu)=m_\tau^2/m_\mu^2$, arising in 2HDMs with natural flavor conservation or flavor alignment. Additionally, the ratio of branching ratios between the LHC most searched decay modes $t b$ and $\tau\nu_\tau$ obeys the relation 
\begin{equation}\label{eq:chargedHBRrel}
\frac{{\rm{BR}}(H^\pm \to t b)}{{\rm{BR}}(H^\pm \to \tau \nu_\tau)} = 3 \times \mathcal{O}\left(\frac{m_c^2}{m_\mu^2}\right) =\mathcal{O}(100),
\end{equation}
valid in the regime of large $\tan\beta$, as opposed to the ratios $3 m_t^2/m_\tau^2\sim 6\times 10^6$, $3 m_b^2/m_\tau^2\sim 1800$, as arising in type I and type II 2HDM, respectively. For this reason, in our model, we expect the $\tau\nu_\tau$ to be relatively more important than the $tb$ mode, if compared to the most studied type I and II 2HDM. We present the results for the branching ratios of the charged Higgs boson in Fig. \ref{fig:BRcharged}, on the left panel as a function of the charged Higgs mass, having fixed $\tan\beta=50$, and on the right panel as a function of $\tan\beta$, having fixed the mass of the charged Higgs to 500 GeV. For both panels, we fix $\cos(\beta-\alpha)=0.05$, in such a way that the $Wh$ charged Higgs partial width is fully determined. Similarly to the neutral heavy Higgs boson, for low values of $\tan\beta$ the largest branching ratios approach the values of a 2HDM of type I, with the $t b$ decay being the dominant one. At large values of $\tan\beta$, instead, the 
decays to second and third generation quarks have comparable branching ratios, and the decay to leptons ($\mu\nu_\mu$ and $\tau\nu_\tau$) are comparable and suppressed by roughly two orders of magnitude, as shown in Eq. (\ref{eq:chargedHBRrel}). Similarly to the neutral Higgs decaying to $WW$ and $ZZ$, at intermediate values of $\tan\beta$, the $Wh$ decay mode can be the dominant one, having fixed $\cos(\beta-\alpha)=0.05$.

\subsection{Production Cross Sections}

\begin{figure*}[tb]
\centering
\includegraphics[width=0.45\textwidth]{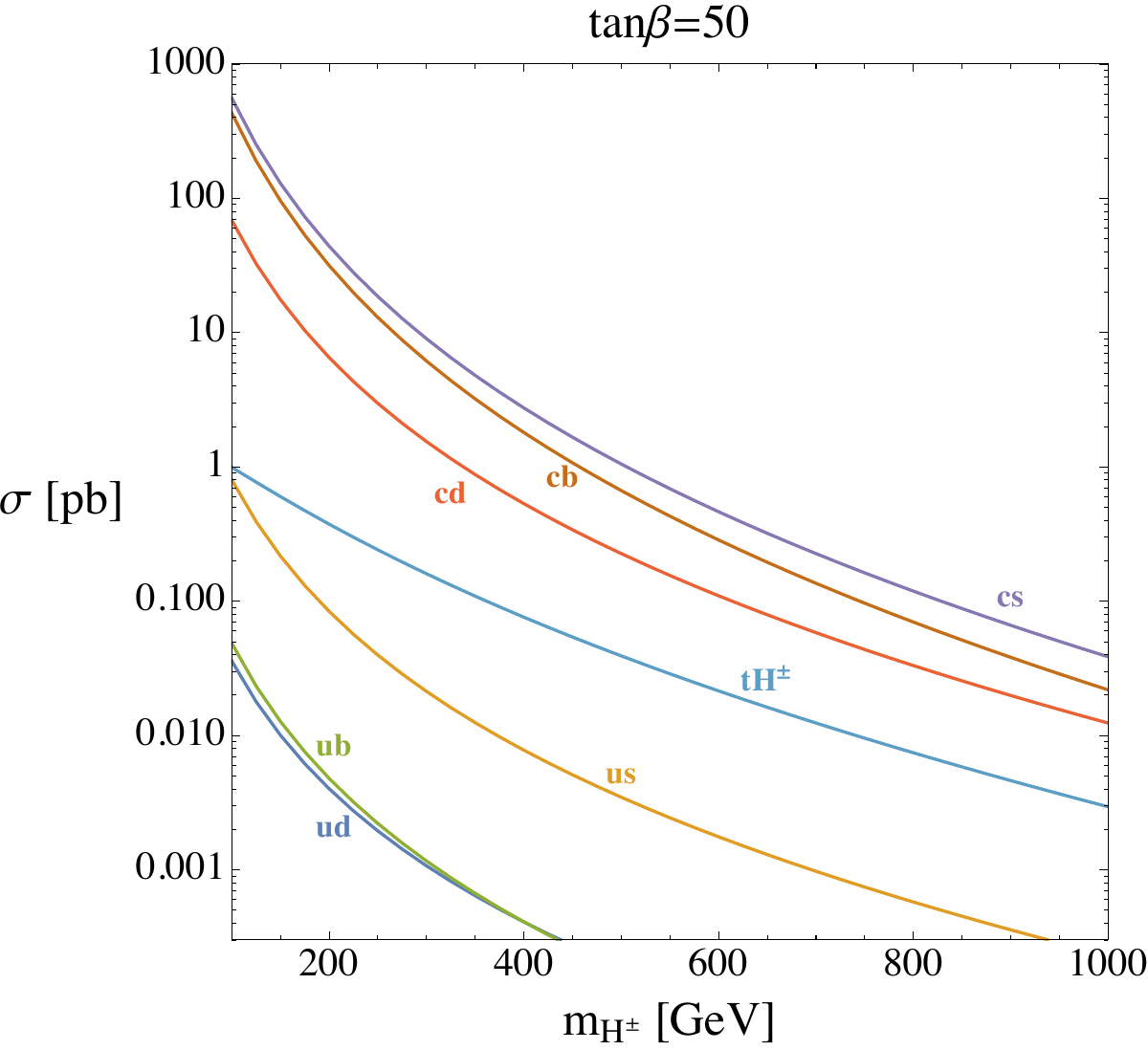} ~~~
\includegraphics[width=0.45\textwidth]{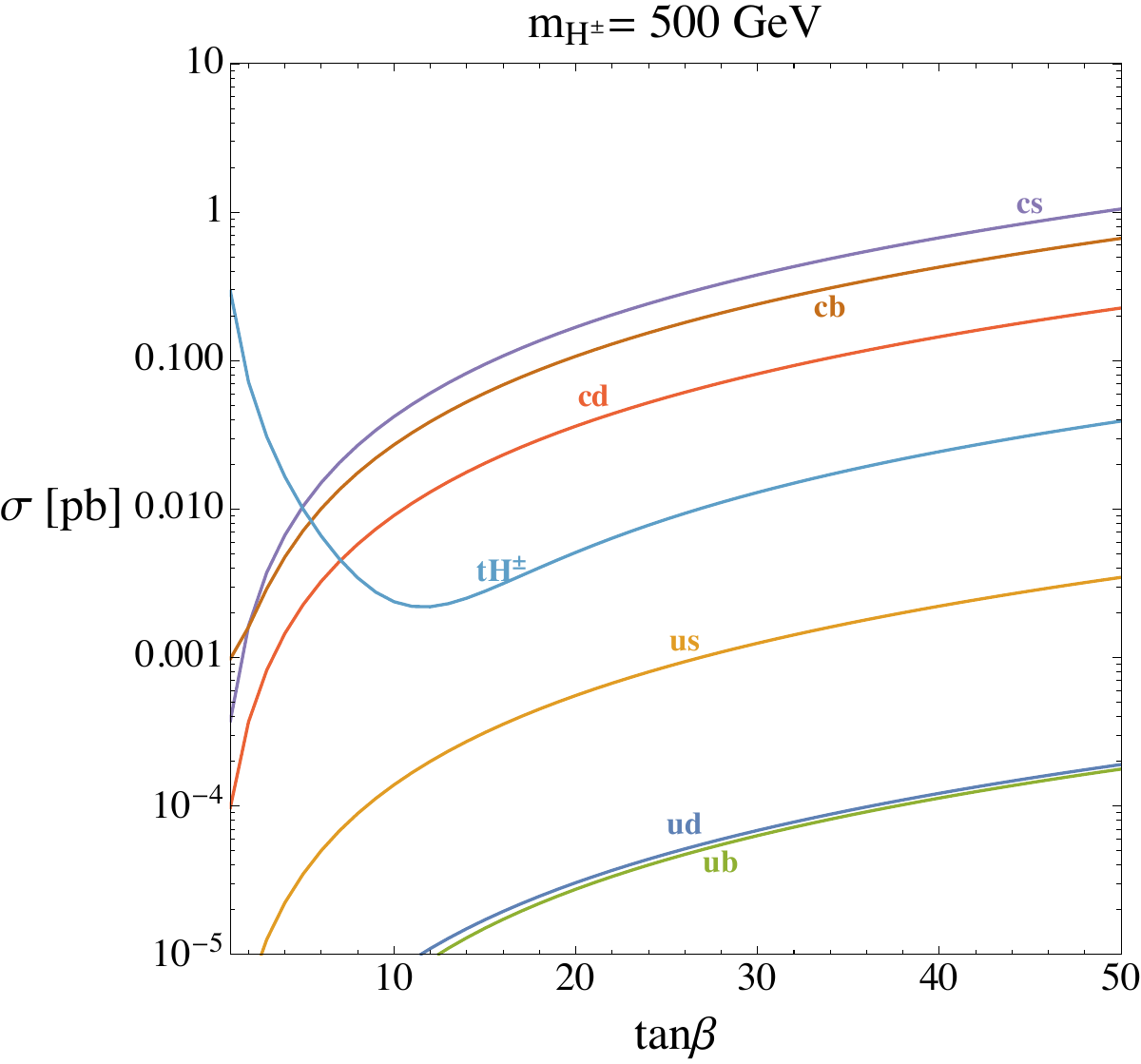}
\caption{Production cross sections of the charged Higgs $H^\pm$ at 13 TeV proton proton collisions as a function of the charged Higgs mass $m_{H^\pm}$ for fixed $\tan\beta = 50$ (left) and as a function of $\tan\beta$ for fixed mass $m_{H^\pm} = 500$~GeV (right). None of these cross sections depend on the value of $\cos(\beta-\alpha)$.}
\label{fig:sigmacharged}
\end{figure*}

In Fig. \ref{fig:sigmacharged}, we show the production cross sections of the charged Higgs at 13 TeV proton-proton collisions as a function of its mass ($m_{H^\pm}>m_t$) for fixed $\tan\beta = 50$ (left) and as a function of $\tan\beta$ for fixed $m_{H^\pm} = 500$ GeV (right). None of these cross sections depend on the value of $\cos(\beta-\alpha)$. For the calculation of these production cross sections, we follow the same procedure as for the neutral Higgs boson. The most interesting features arise at moderate and sizable values of $\tan\beta$, as at small values of $\tan\beta$ the main production cross section comes from the $tH^\pm$ associated process, as predicted by the most studied type I and type II 2HDMs. At larger values of $\tan\beta$, the production cross sections from $cs$, $cb$, $cd$ are also very important and can even dominate over $tH^\pm$. These production cross sections are all of the same order and their exact size depends strongly on the specific values of the $m^\prime$ parameters. 
Similarly to the neutral Higgs, the inclusive cross section is at the level of few $\times~100$ fb over a broad range of masses. In the figure, we do not show the cross section for the associated production $pp\to H^\pm h$ since it is typically below the fb level for $\cos(\beta-\alpha)=0.05$.

For $m_{H^\pm}<m_t$, the charged Higgs is mainly produced from the top decay modes $t\to H^\pm b$ and $t \to H^\pm s$. The branching ratios for these processes are at around few $\%$ for $\tan\beta=50$.

\section{Experimental Sensitivities and New Signatures} \label{sec:signals}

\begin{figure*}[tb]
\centering
\includegraphics[width=0.45\textwidth]{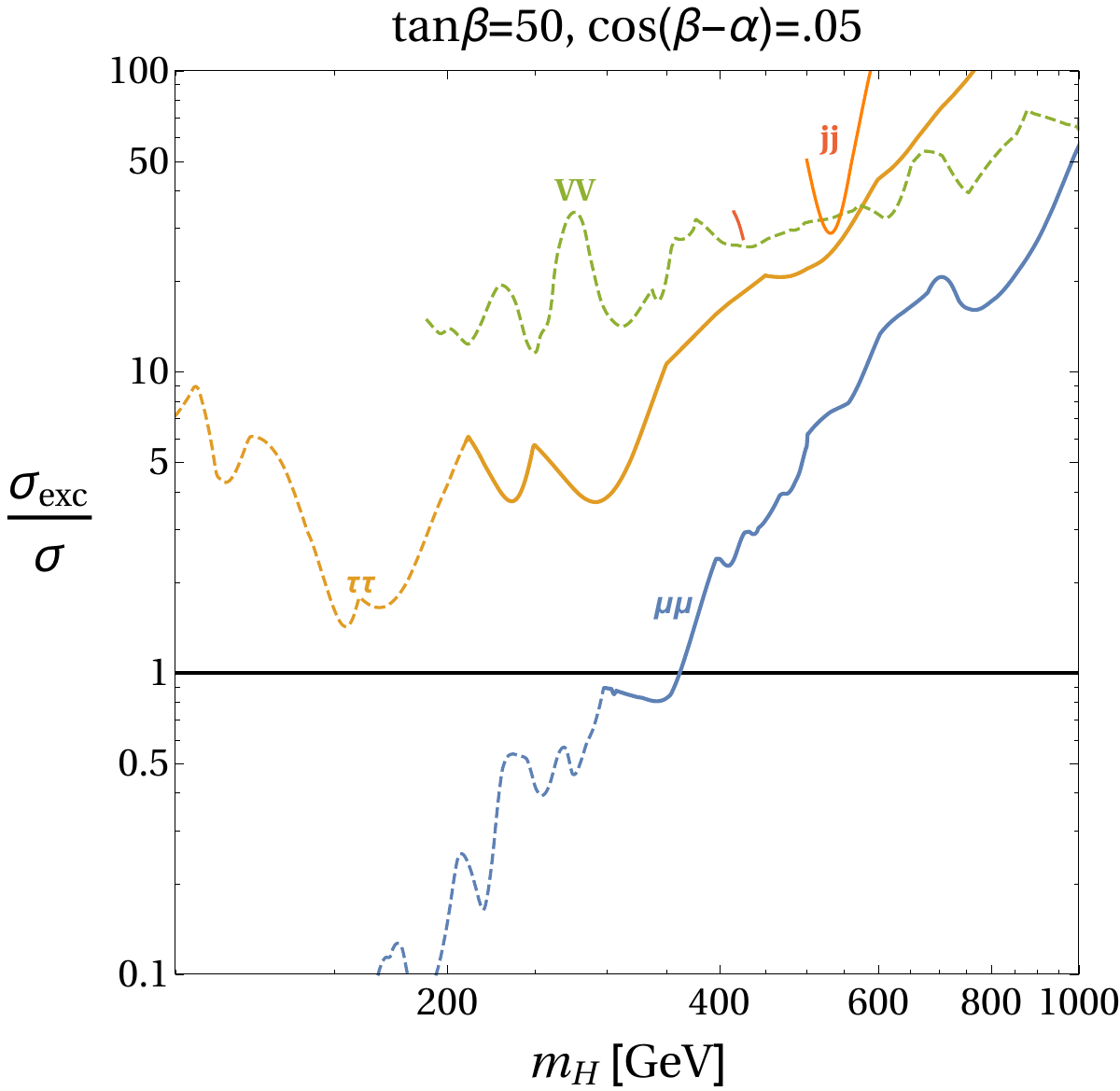} ~~~
\includegraphics[width=0.5\textwidth]{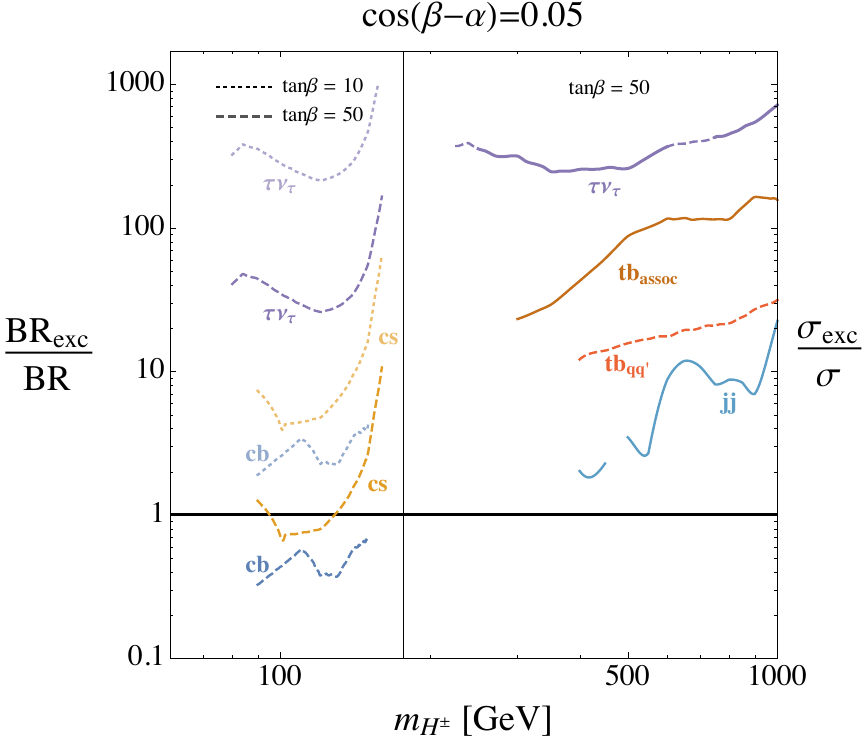}
\caption{Experimental exclusion limits normalized to the predicted cross sections for the heavy scalar boson (left) and for the charged Higgs (right) as a function of the corresponding Higgs mass. We set $\tan\beta = 50$ and $\cos(\beta-\alpha) = 0.05$. Shown are the currently most strigent constraints coming from searches for $\tau^+\tau^-$, $ZZ/WW$, $jj$, and $\mu^+\mu^-$ final states (neutral scalar) and $cb$, $cs$, $\tau\nu$, $tb$, and $jj$ final states (charged Higgs). The solid (dashed) curves correspond to 13~(8)~TeV analyses.}
\label{fig:exp}
\end{figure*}

After discussing the branching ratios and production cross sections separately for the neutral and charged Higgs bosons, we confront our model with existing searches for additional Higgs bosons at the LHC. Searches for neutral Higgses have been performed at 8~TeV and 13~TeV in a variety of channels including
\begin{itemize}
 \item[(i)] $H\to ZZ$ and $H \to WW$~\cite{Khachatryan:2015cwa,Aad:2015kna,Aad:2015agg,CMS:2016rqf,atlas_ww_13TeV,CMS:2016jpd,ATLAS:2016kjy,CMS:2016ilx,ATLAS:2016oum},
 \item[(ii)] $A\to Zh$~\cite{Aad:2015wra,atlas_azh_13TeV,Khachatryan:2015lba,Khachatryan:2015tha},
 \item[(iii)] $A/H \to \tau^+\tau^-$~\cite{Khachatryan:2014wca,Aad:2014vgg,atlas_tautau_13TeV,CMS:2016pkt,ATLAS:2016fpj},
  \item[(iv)] $A/H \to \mu^+\mu^-$~\cite{CMS:2015ooa},
\item[(v)] $A/H \to t\bar t$~\cite{ATLAS:2016pyq}.
\end{itemize}
Moreover, we also take into account generic searches for
\begin{itemize}
 \item[(vi)] di-muon resonances~\cite{Aad:2014cka,Khachatryan:2014fba,atlas_dimuon_13TeV,CMS:2015nhc,CMS:2016abv,ATLAS:2016cyf},
 \item[(vii)] di-jet resonances~\cite{CMS:2015neg,ATLAS:2016xiv,Khachatryan:2016ecr,ATLAS:2016bvn,ATLAS:2016jcu,CMS:2016jog},
\end{itemize}
which, as we will discuss, have interesting sensitivities to our parameter space.

On the left panel of Fig.~\ref{fig:exp} we show the ratio of currently excluded cross section over the cross section predicted in our model as a function of the scalar Higgs mass $m_H$. A ratio smaller than 1 indicates exclusion.
In the plots we set $\tan\beta = 50$ and $\cos(\beta-\alpha) = 0.05$.
For a given Higgs mass we show the strongest constraint of a specific category of final states ($\tau^+\tau^-$, $ZZ/WW$, $jj$, $\mu^+\mu^-$). The solid (dashed) lines indicate 13~(8)~TeV analyses. 

The 8~TeV inclusive search for $H \to \mu^+\mu^-$~\cite{CMS:2015ooa} is the most sensitive at low masses. At higher masses $m_H \gtrsim 300$~GeV, the 13~TeV searches for di-muon resonances turn out to be most sensitive. In comparing the excluded cross sections with our model predictions we add up gluon fusion and production from charm initial states, since we do not expect that the signal efficiencies differ significantly for these production modes.
We find that the $H \to \mu^+\mu^-$ searches exclude the heavy scalar with mass $m_H \lesssim 360$~GeV for $\tan\beta = 50$. 
For lower $\tan\beta$, this constraint becomes weaker, due to the smaller production cross sections, and it does not extend the LEP bound for $\tan\beta \lesssim 12$.

Searches for $H \to \tau^+\tau^-$ give strong constraints on 2HDMs of type~II in the large $\tan\beta$ regime. In our model, on the other hand, the small branching ratio of $H \to \tau^+\tau^-$ renders these searches less relevant. Even for $\tan\beta = 50$, we find that current experimental sensitivities do not yet allow to probe the heavy scalar using this channel. 

Searches for $H \to ZZ$ currently constrain cross sections that are approximately one order of magnitude larger than those of the benchmark shown in Fig.~\ref{fig:exp}. These searches can become relevant for moderate $\tan\beta$ and larger $\cos(\beta- \alpha)$.
Searches for $H \to WW$ are generically less sensitive as compared to $H \to ZZ$. The corresponding channel for the pseudo-scalar, which has similar sensitivity, is $A\to Zh$.

Given the large branching ratio $H \to c \bar c$ (see Fig. \ref{fig:BR}) also searches for light di-jet resonances might be interesting. The ATLAS di-jet search performed with 3.4~fb$^{-1}$ 13~TeV data \cite{ATLAS:2016xiv} using a trigger-object level analysis sets a constraint on the model $\sim 1$ order of magnitude more stringent than the 8 TeV analyses performed by CMS with data scouting \cite{CMS:2015neg,Khachatryan:2016ecr}, reaching the best sensitivity to our model for masses at around $550$ GeV. We also checked the performance of the analyses \cite{ATLAS:2016bvn,ATLAS:2016jcu,CMS:2016jog} in testing our model. These CMS and ATLAS searches focus on the production of a (light) di-jet resonance in association with a boosted photon or jet. Due to the very high $p_T$ threshold required for this additional object, these searches are less sensitive to our scenario, if compared to the trigger-object level analysis~\cite{ATLAS:2016xiv}.
the corresponding cross section predicted by our model. 
As we can see from the left panel of Fig. \ref{fig:exp}, the di-jet constraints are comparable (or even stronger, for some values of $m_H$) to the constraints from the most studied $H\to\tau^+\tau^-$ searches. 

Finally, in the figure we do not show the constraints from $A/H \to t\bar t$~\cite{ATLAS:2016pyq}, as they are very weak. This is due to the interference of the signal with the SM $t\bar t$ continuum~\cite{Craig:2015jba,Jung:2015gta,Gori:2016zto,Carena:2016npr}.

\bigskip
For the charged Higgs we consider searches for
\begin{itemize}
 \item[(i)] $(t)H^\pm\to \tau\nu$~\cite{Aaboud:2016dig,Khachatryan:2015qxa,ATLAS:2016grc, Aad:2014kga}, for both the charged Higgs mass below and above the top mass.
 \item[(ii)] $H^\pm\to t b$: ~\cite{Aad:2015typ}, both for $pp\to tH^\pm$ and $qq^\prime\to H^\pm$ production;~\cite{Khachatryan:2015qxa,ATLAS:2016qiq}  for $pp\to (b)tH^\pm$,
 \item[(iii)] $H^\pm \to c s$~\cite{Khachatryan:2015uua}, for $m_{H^\pm}<m_t$,
 \item [(iv)] $H^\pm \to c b$~\cite{CMS:2016qoa} for $m_{H^\pm}<m_t$,
 \item[(v)] $H^\pm \to Wh$~\cite{Khachatryan:2015bma,Khachatryan:2016yji,CMS:2016dzw,ATLASWh,Aaboud:2016lwx},
 \item[(vi)] $H^\pm \to \mu \nu_\mu$~\cite{ATLAS:2016ecs,CMS:2015kjy},
 \item[(vii)] generic searches for di-jet resonances~\cite{CMS:2015neg,ATLAS:2016xiv,Khachatryan:2016ecr,ATLAS:2016bvn,ATLAS:2016jcu,CMS:2016jog}.
\end{itemize}

In the right panel of Fig. \ref{fig:exp}, we only show the bounds from $H^\pm\to\tau\nu$, $H^\pm\to tb$, $H^\pm\to cs$, $H^\pm\to cb$, and searches for di-jet resonances. We do not show the bound from the $Wh$ decay, as these searches are performed only for very heavy resonances $m_{H^\pm}\gtrsim 800$ GeV and lead only to very weak constraints on the parameter space of our model. Also bounds from $\mu \nu_\mu$ searches are not shown. They do not lead to interesting constraints since the $H^\pm\to\mu\nu_\mu$ branching ratio, despite being enhanced compared to 2HDMs of type I or II, is not large enough in our model.

Below the top mass, the most stringent constraint comes from the $cb$ search \cite{CMS:2016qoa} performed with the full 8 TeV data set. This is followed by the 8 TeV $cs$ search \cite{Khachatryan:2015uua}\footnote{The bounds we are presenting in the figure for $m_{H^\pm}<m_t$ are a conservative estimates, since they do not keep into account the possible pollution of events coming from the process $t\to s H^\pm$ with a strange quark mis-tagged to be a b-quark.}. 
For $\tan\beta=50$ charged Higgs masses above the LEP bound and below $\sim 160$ GeV are fully probed by these searches (see dashed lines in the right panel of Fig. \ref{fig:exp} for $m_{H^\pm}<m_t$). However, the bound gets significantly weaker for intermediate values of $\tan\beta$, as the charged Higgs production cross section gets smaller: as shown by the dotted lines obtained for $\tan\beta=10$, the entire mass range below the top mass opens up. For even smaller values of $\tan\beta$ the charged Higgs production increases again, leading to stronger bounds, if compared to $\tan\beta=10$.

Above the top mass, the most important constraint comes from the search of $tb$ resonances, that are, however, not able to set any bound on our model. Particularly, the process $qq^\prime\to H^\pm\to tb$~\cite{Aad:2015typ} (denoted by $tb_{qq^\prime}$ in the figure) is presently probing cross sections up to $\sim 10$ bigger than the cross sections predicted by our model for $\tan\beta=50$. The 13 TeV search for $pp\to (b) tH^\pm, H^\pm\to tb$~\cite{ATLAS:2016qiq} offers only weaker bounds, due to the production cross section for $tH^\pm$ being more than one order of magnitude smaller than the corresponding $qq^\prime\to H^\pm$ (see right panel of Fig. \ref{fig:sigmacharged}).
Searches for di-jet resonances have sensitivities that are comparable to the search of $tb$ resonances. 
To estimate the di-jet signal from the charged Higgs we take into account the charged Higgs production from $cs$, $cb$, and $cd$ initial states and all charged Higgs branching ratios into quarks except those including a top quark.  
The highest sensitivity comes from the 13~TeV ATLAS search~\cite{ATLAS:2016xiv} and is shown in the plot by the line denoted by $jj$.

\begin{figure*}[tb]
\centering
\includegraphics[width=0.45\textwidth]{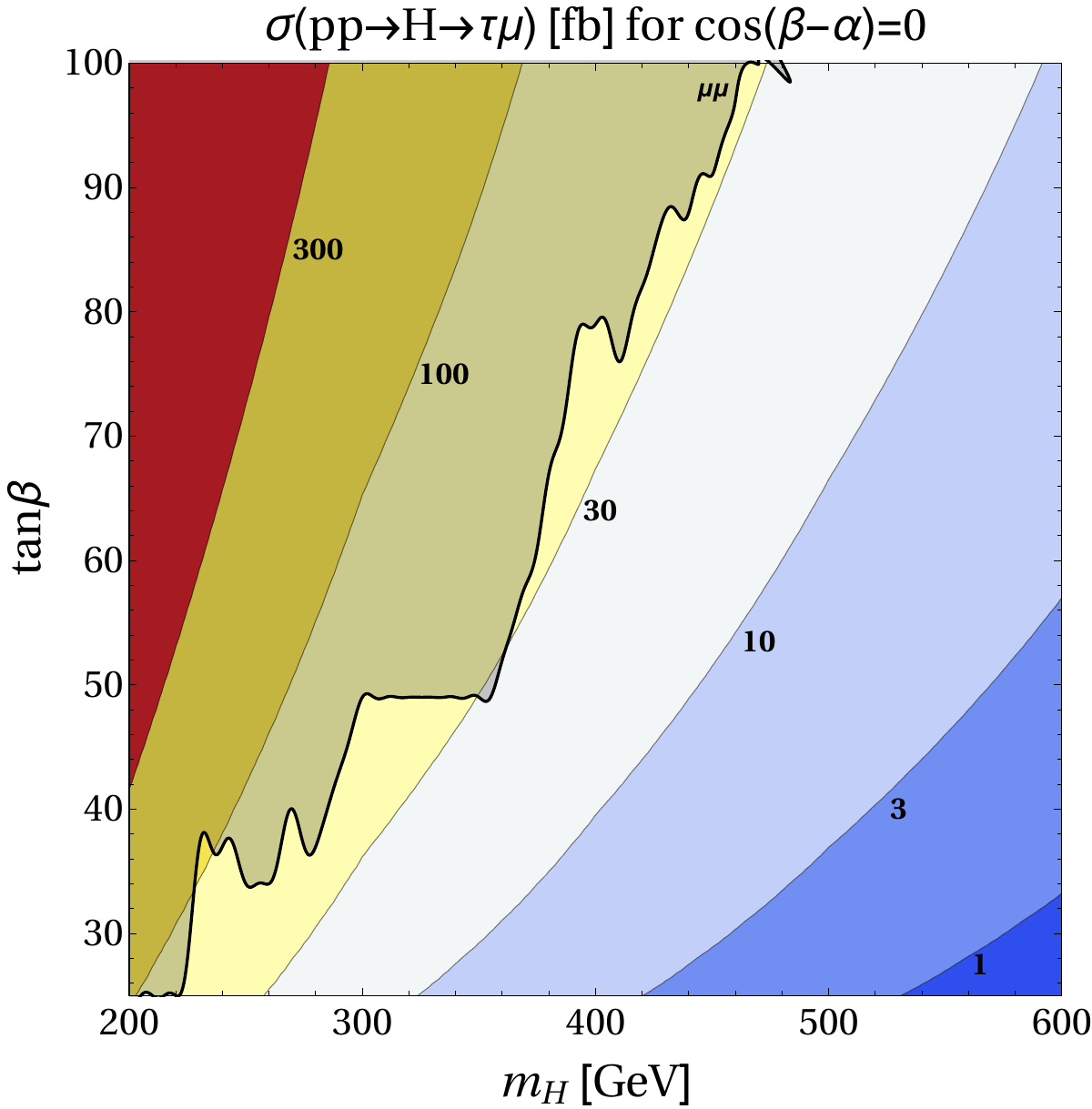} ~~~
\includegraphics[width=0.45\textwidth]{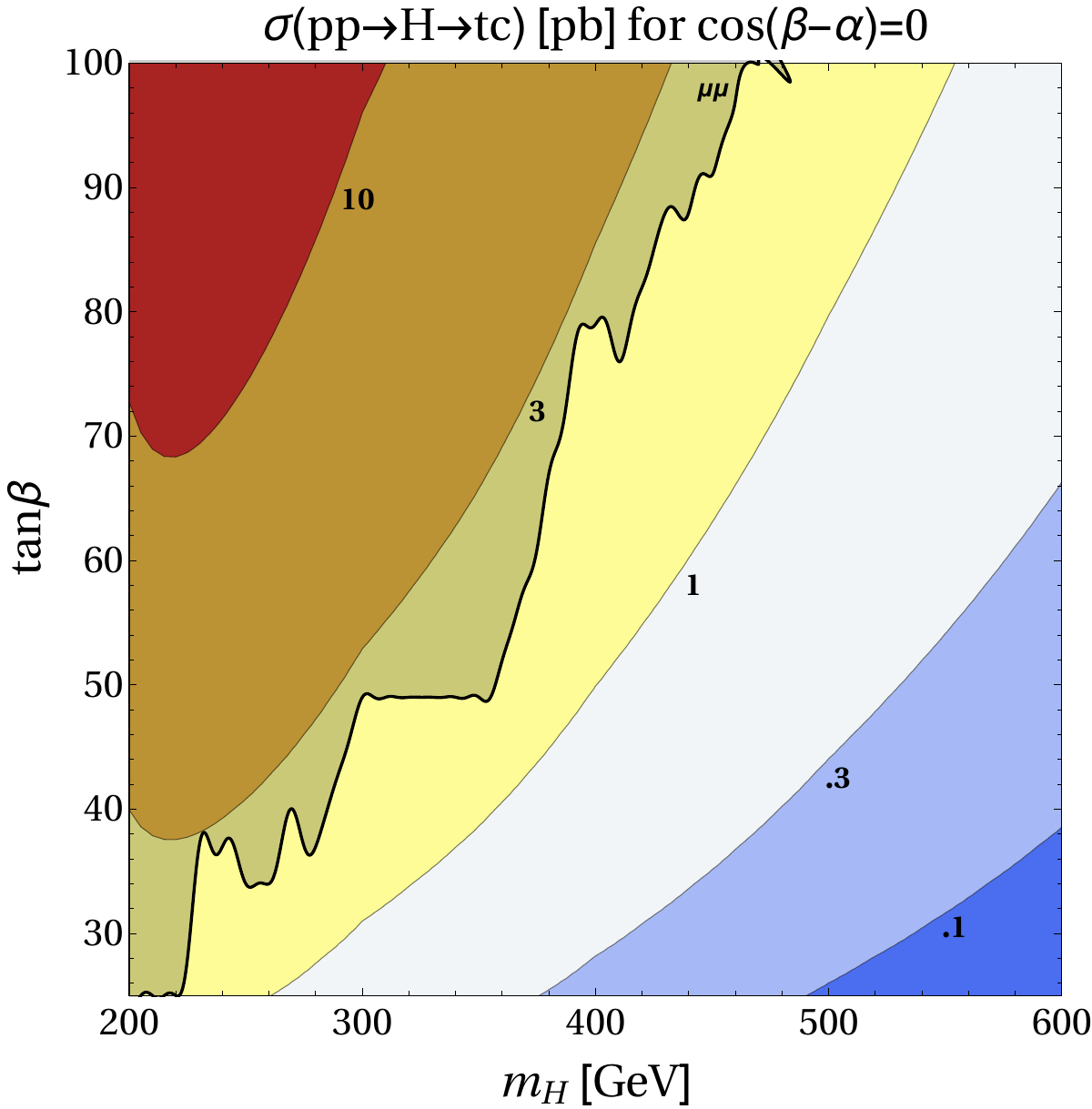}\\[16pt]
\includegraphics[width=0.45\textwidth]{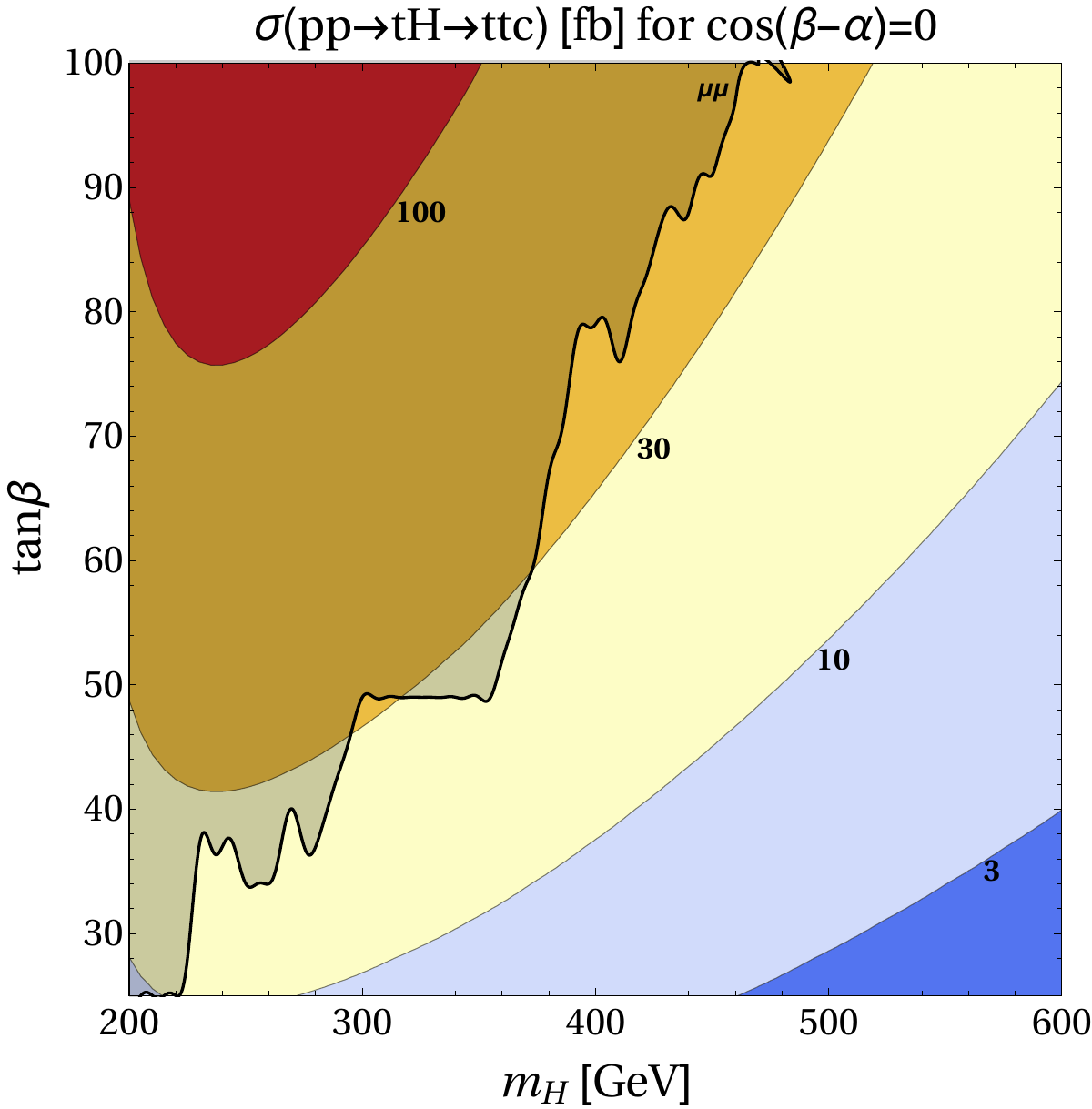} ~~~
\includegraphics[width=0.45\textwidth]{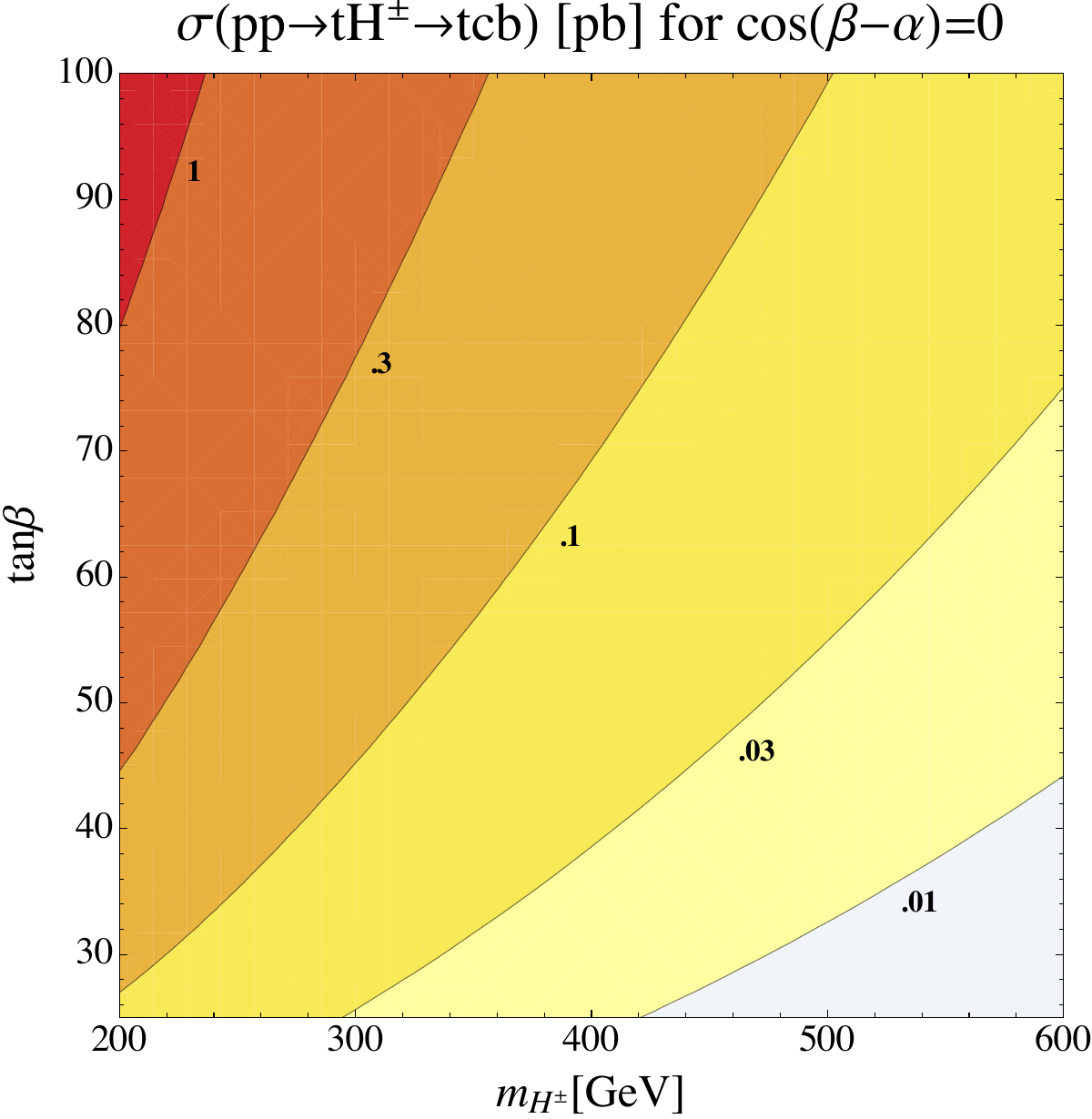}
\caption{Production cross section times branching ratio for the processes $pp \to H \to \tau\mu$ (upper left), $pp \to H \to tc$ (upper right) and $pp \to t H , H\to tc$ (lower left) at 13~TeV in the $m_H$ vs. $\tan\beta$ plane in the decoupling or alignment limit, $\cos(\beta-\alpha)=0$. The gray shaded region is excluded by existing searches for di-muon resonances. Lower right panel: Production cross section times branching ratio for the process $pp \to t H^\pm, H^\pm cb$ at 13~TeV in the $m_{H^\pm}$ vs. $\tan\beta$ plane in the decoupling or alignment limit.}
\label{fig:signatures}
\end{figure*}

\bigskip
Our model also predicts a set of novel signatures that can be searched for at the LHC.
Interesting signatures include flavor-violating neutral Higgs decays $pp \to H/A \to \tau \mu$ and $pp \to H/A \to tc$ and multi-top final states $pp \to t H/A \to ttc$.
Cross sections for the processes involving the scalar, $H$, are shown in the $m_H - \tan\beta$ plane in the upper and lower left panels of Fig.~\ref{fig:signatures}, having fixed $\cos(\beta-\alpha)=0$.

Compared to a 2HDM type II, a much larger region of the $m_H - \tan\beta$ plane is not yet probed by existing searches. In a 2HDM type II, searches for $H/A \to \tau^+\tau^-$ are sensitive to neutral Higgs bosons with masses of 300-400 GeV as long as $\tan\beta \gtrsim 15$ \cite{Khachatryan:2014wca,Aad:2014vgg,atlas_tautau_13TeV,CMS:2016pkt,ATLAS:2016fpj}. For $\tan\beta \gtrsim 50$, neutral Higgs bosons above 1~TeV can be probed.
In our setup, the sensitivity of $H/A \to \tau^+\tau^-$ searches is weak. As discussed above, the most important constraints can be derived from di-muon resonance searches that are sensitive to neutral Higgs bosons of $\sim 290$~GeV for $\tan\beta \sim 50$. The parameter space that is excluded by current di-muon resonance searches is shaded in gray in the upper and lower left plots of Fig.~\ref{fig:signatures}. 

In the allowed parameter space, the $pp \to H \to \tau \mu$ cross section can be several 10s of fb up to 100 fb.
The $pp \to H \to tc$ cross section can be as large as few pb.
Finally, the $pp \to t H \to ttc$ cross section can reach $\sim 30$~fb in the shown scenario. Cross sections that are larger by a factor of few are easily possible by modifying the free parameters $m_{tc}^\prime$ and $m_{ct}^\prime$ that control the size of the $H tc$ coupling. 
Interestingly enough, generically one half of this cross section corresponds to same sign tops $pp \to t H \to tt \bar c$ or $pp \to \bar t H \to \bar t \bar t c$, providing a very distinct signature of this model. 

As shown in the right panel of Fig. \ref{fig:exp}, the parameter space of the charged Higgs above the top mass is completely un-constrained by the current LHC analyses, even at large values of $\tan\beta$ ($=50$ in the figure). However, notice that there are indirect constraints from the neutral Higgses, because their mass cannot differ too much from the charged Higgs mass.
It will be very interesting to design new searches to look for our charged Higgs in the coming years of the LHC.
In particular, the cross section for $pp\to t H^\pm$ with $H^\pm\to cb$ can be at the few hundreds fb - pb level in a large range of parameters for $m_{H^\pm}>m_t$ (see lower right panel of Fig. \ref{fig:signatures}). Additionally, the cross section for the flavor conserving signature $pp\to t H^\pm$ with $H^\pm\to cs$ has similar values. This offers a unique opportunity to look for a di-jet resonance (eventually with a b-tag) produced in association with a top quark. Finally, our model also predicts the novel interesting signature $pp\to t H^\pm$ with $H^\pm\to \mu^\pm\nu_\mu$, but the cross section is at the fb level even for $\tan\beta=50$. Therefore, it will be likely more difficult to probe our charged Higgs using this signature.

\section{Summary} \label{sec:conclusions}

We discussed the distinct collider phenomenology of a class of 2HDMs in which the 125 GeV Higgs is mainly responsible for the masses of the weak gauge bosons and of the third generation fermions, while the second Higgs doublet provides mass for the lighter fermion flavors. This model is particularly well motivated in view of our ignorance concerning the coupling of the 125 GeV Higgs to first two generation quarks and leptons.

The 125~GeV Higgs has modified couplings to SM fermions that qualitatively deviate from the couplings in 2HDMs with natural flavor conservation, minimal flavor violation, or flavor alignment.
While the 125~GeV Higgs couplings to the third generation fermions behave as in a 2HDM type I and are close to their SM values, all couplings to second and first generation fermions can be easily modified by $\mathcal{O}(1)$. We find that the searches for $h \to \mu^+\mu^-$ provide the strongest constraints on deviations from the decoupling limit $\cos(\beta-\alpha) = 0$ for moderate and large values of $\tan\beta$. 
The framework predicts generically a $\mathcal{O}(0.1\%)$ flavor-violating branching ratio $h \to \tau\mu$.

The heavy neutral Higgs bosons, $H$ and $A$, have a very distinct phenomenology. They have couplings to second and first generation fermions that are enhanced by $\tan\beta$, while their couplings to the third generation are suppressed.
For large $\tan\beta$, we generically find that the dominant decay modes are into $c\bar c$, $t \bar t$, and $c t$ with branching ratios that are comparable in size. Branching ratios for decays into final states involving gauge bosons ($H \to WW/ZZ$ and $A \to Zh$) can be sizable for moderate values of $\tan\beta$. 
Decays into $\mu^+\mu^-$, $\tau^+ \tau^-$, and $\tau \mu_\tau$ are typically also comparable and the corresponding branching ratios can reach the \% level. 
The most important production modes are gluon fusion and production from charm initial states. For large $\tan\beta$, the cross section from charm can be several hundreds of fb for a Higgs mass of 500~GeV.

The charged Higgs boson is mainly produced by second and third generation quark fusion, as well as in association with a top. Its decays are interestingly different from the decays arising in type I and II 2HDMs, as they are dominated by flavor-violating $cb, ts$ decays and by decays to second generation $cs$. Also the hierarchy between the decay rate into $\mu\nu_\mu$ and into $\tau\nu_\tau$ is not the same as in 2HDMs with natural flavor conservation or flavor alignment, as the muon decay is parametrically enhanced. This results in weak bounds from the LHC most searched-for signatures, $tb$ and $\tau\nu_\tau$. 

Due to the non-standard branching ratios and production modes of $H$, $A$, and $H^\pm$, the standard searches for heavy Higgs bosons are not necessarily the most sensitive probes of our extended scalar sector.
We find that, currently, the searches for low mass di-muon resonances place the most stringent constraints on the model.
Also searches for low mass di-jet resonances might probe interesting parameter space in the future. Interesting novel signatures include heavy neutral Higgs bosons decaying in a flavor-violating way, e.g. $pp \to H/A \to \tau \mu$ or $pp \to H/A \to t c$, as well as final states with same sign tops $pp \to tH  \to tt \bar c$ or $pp \to \bar t H  \to \bar t \bar t c$. For the charged Higgs, it will be very interesting to perform searches for $cb$ and $cs$ resonances with mass above the top threshold, produced in association with a top quark.

\section*{Acknowledgements}

We thank Bill Murray for useful discussions.
WA and SG acknowledge financial support by the University of Cincinnati. WA and SG thank the Aspen Center for Physics for hospitality during the completion of this work. The Aspen Center for Physics is supported by
National Science Foundation grant PHY-1066293. WA is grateful to the Mainz Institute for Theoretical Physics (MITP) for its hospitality and its partial support during final stages of this work. S.G. is grateful for the hospitality of the Kavli Institute for Theoretical Physics in Santa Barbara, CA, supported in part by the National Science Foundation under Grant No. NSF PHY11-25915. The work of JE was partially supported by a Mary J. Hanna Fellowship through the Department of Physics at University of Cincinnati, and also by the U.S. Department of Energy, Office of Science, Office of Workforce Development for Teachers and Scientists, Office of Science Graduate Student Research (SCGSR) program. The SCGSR program is administered by the Oak Ridge Institute for Science and Education for the DOE under contract number DE-SC0014664. The work of ML was partially supported by a Graduate Student Research Fellowship of the University Research Council at the University of Cincinnati and by the Mary J. Hanna Fellowship 
through the Physics Department at the University of Cincinnati. MM work was supported in part by DOE grant DE-SC0011784 and in part by NSF grant PHY-1151392.


\newpage{\pagestyle{empty}\cleardoublepage}

\end{document}